\documentclass[twocolumn]{aastex631}

\newcommand{\be}{\begin{displaymath}}
\newcommand{\ee}{\end{displaymath}}
\newcommand{\bea}{\begin{eqnarray*}}
\newcommand{\eea}{\end{eqnarray*}}
\newcommand{\snr}{SNR$_1$}

\shorttitle{{\it Gaia} Exoplanet Forecasts}
\shortauthors{Lammers \& Winn}

\graphicspath{{./}{}}
\usepackage{amsmath}
\usepackage{comment}

\begin{document}

\title{On the Exoplanet Yield of {\it Gaia} Astrometry}


\author[0000-0001-9985-0643]{Caleb Lammers}
\affiliation{Department of Astrophysical Sciences, Princeton University, 4 Ivy Lane, Princeton, NJ 08544, USA}

\author[0000-0002-4265-047X]{Joshua N.\ Winn}
\affiliation{Department of Astrophysical Sciences, Princeton University, 4 Ivy Lane, Princeton, NJ 08544, USA}

\begin{abstract}

We re-examine the expected yield of {\it Gaia} astrometric planet detections using updated models for giant-planet occurrence, the local stellar population, and {\it Gaia}'s demonstrated astrometric precision. Our analysis combines a semi-analytic model that clarifies key scaling relations with more realistic Monte Carlo simulations. We predict $7{,}500\,{\pm}\,2{,}100$ planet discoveries in the 5-year dataset (DR4) and $120{,}000\,{\pm}\,22{,}000$ over the full 10-year mission (DR5), with the dominant error arising from uncertainties in giant-planet occurrence. We evaluate the sensitivity of these forecasts to the detection threshold and the desired precision for measurements of planet masses and orbital parameters. Roughly $1{,}900\,{\pm}\,540$ planets in DR4 and $38{,}000\,{\pm}\,7{,}300$ planets in DR5 should have masses and orbital periods determined to better than $20$\%. Most detections will be super-Jupiters ($3$\,--\,$13\,M_{\rm J}$) on $2$\,--\,$5$\,AU orbits around GKM-type stars ($0.4$\,--\,$1.3\,M_\odot$) within $500$~pc. Unresolved binary stars will lead to spurious planet detections, but we estimate that genuine planets will outnumber them by a factor of $5$ or more. An exception is planets around M-dwarfs with $a\,{<}\,1$\,AU, for which the false-positive rate is expected to be about $50$\%. To support community preparation for upcoming data releases, we provide mock catalogs of {\it Gaia} exoplanets and planet-impostor binaries.\footnote{\url{https://github.com/CalebLammers/GaiaForecasts}}

\end{abstract}

\section{Introduction}
\label{sec:intro}

The astrometric method for planet detection requires monitoring the position of a star on the celestial sphere with a precision that has historically been difficult to achieve. Many exoplanet discoveries reported based on ground-based optical astrometry have proven spurious \citep[see, e.g.,][]{Hunter1943, Holmberg1938, Reuyl&Holmberg1943, Strand1943, Lippincott1960, vandeKamp1963, Hershey&Lippincott1982, Gatewood1996, Pravdo&Shaklan2009}, although there remain a few viable candidates \citep{Muterspaugh+2010, GravityCollaboration2024}. Very-long-baseline radio interferometry can in principle deliver the necessary precision, but is applicable to few stars, and has yielded only one planet to date \citep{Curiel2022}.

The {\it Gaia} mission promises a new beginning for astrometric planet discovery. {\it Gaia} provides space-based positional measurements for over a billion stars across the entire sky \citep{GaiaCollaboration2016}. For nearby stars, the precision is sufficient to detect giant planets, and {\it Gaia}'s planet discovery potential was recognized early in mission planning. \citet{Perryman2001} anticipated about $30{,}000$ planet detections in the five-year mission, although planet occurrence rates and {\it Gaia}'s ultimate precision were uncertain at that time (see also \citealt{Lattanzi2000, Quist2001, Sozzetti2001}). \citet{Casertano2008} refined the estimate to ${\sim}\,8{,}000$ giant planets using the Besançon stellar population model \citep{Robin2003} and injection-recovery tests, but they restricted the sample to FGK dwarfs with $V\,{<}\,13$ within $200$~parsecs and assumed astrometric uncertainties several times smaller than have been realized. Focusing on nearby M-dwarfs, \citet{Sozzetti2014} predicted that there would be roughly $2{,}600$ detections.

\begin{figure*}
\centering
\includegraphics[width=0.95\textwidth]{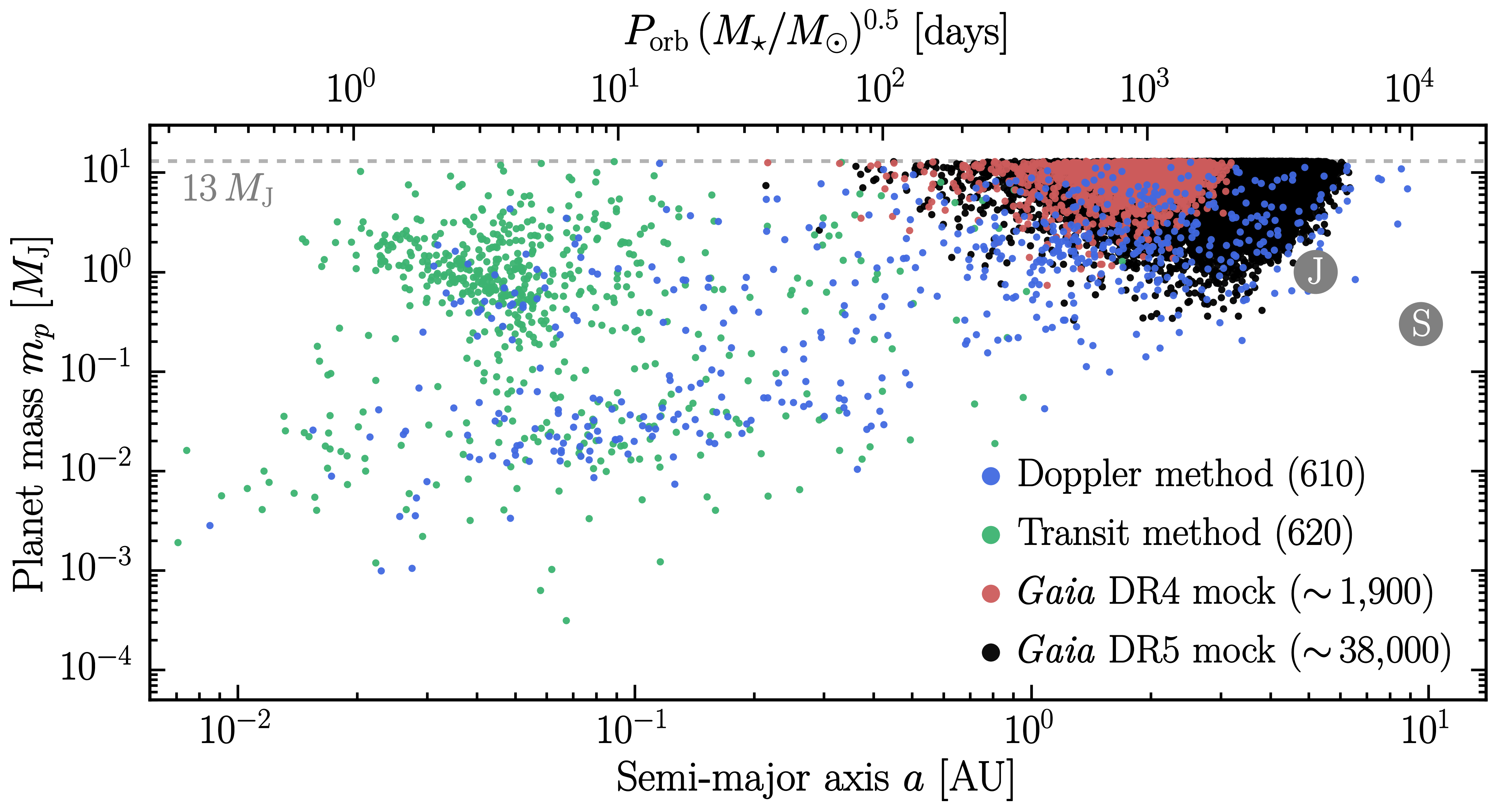}
\caption{Masses and semi-major axes of known planets and simulated {\it Gaia} planets. In both cases, we have restricted the sample to planets with $P_\mathrm{orb}$ and $m_p$ constrained to within $20$\% (i.e., $P_\mathrm{84th}/P_\mathrm{16th}\,{<}\,1.2$ and $m_\mathrm{84th}/m_\mathrm{16th}\,{<}\,1.2$). Blue and green points are real planets discovered with the Doppler and transit methods. Red and black points are from our mock {\it Gaia} planet catalogs for DR4 and DR5 (see Section~\ref{sec:mock_catalog}). For comparison, the values of Jupiter and Saturn are highlighted with gray points labeled ``J'' and ``S.'' The {\it Gaia} planets are primarily super-Jupiters on several-AU orbits. About $100$ such planets are currently known; {\it Gaia} promises to expand this sample to thousands in DR4 and tens of thousands in DR5.}
\label{fig:parameter_space}
\end{figure*}

The most comprehensive prior forecast was by \citet{Perryman2014}, who modeled the local stellar population with TRILEGAL \citep{Girardi2005} and adopted more realistic astrometric uncertainties. They predicted that {\it Gaia} would detect ${\sim}\,21{,}000$ planets in its initial 5-year mission and ${\sim}\,70{,}000$ planets if extended to $10$~years. The 5-year mission planets were predominantly massive giants ($1$\,--\,$15\,M_\mathrm{J}$) on $0.5$\,--\,$5$\,AU orbits around stars within about $600$\,pc.

The next {\it Gaia} data release, scheduled for December 2026, will for the first time provide extensive time-series astrometry spanning $5.5$ years.\footnote{According to \url{https://www.cosmos.esa.int/web/gaia/dr4} (accessed September 20th, 2025).} We revisited {\it Gaia}'s planet yield for four main reasons. First, our understanding of giant-planet occurrence, the local stellar population, and {\it Gaia}'s astrometric precision has improved substantially over the past decade. Second, beyond numerical simulations, we sought an analytic framework to clarify how yields scale with stellar
properties, planet properties, detection thresholds, and observing baseline. Third, we wanted to quantify the burden of spurious detections caused by unresolved stellar binaries, which appear to be frequent in the earliest catalog of {\it Gaia} astrometric planet candidates \citep{Marcussen2023, Stefansson2025}. Finally, we wished to provide the community with mock catalogs of {\it Gaia} planets and planet impostors to guide expectations, help plan follow-up observations, and facilitate eventual comparisons between predictions and observations. Figure~\ref{fig:parameter_space} previews our mock catalogs and illustrates {\it Gaia}'s discovery potential.

We have organized this paper as follows. Section~\ref{sec:analytic_model} presents a semi-analytic model for {\it Gaia}'s exoplanet yield. Section~\ref{sec:astrom_data} describes more realistic simulations in which orbits are fitted to synthetic time-series astrometry, and Section~\ref{sec:mock_catalog} presents the resulting mock planet catalogs. Section~\ref{sec:binary_stars} examines false positives from unresolved binaries. Finally, Section~\ref{sec:discussion} summarizes the results, compares them with previous projections, and notes some limitations. Throughout, we distinguish predictions for Data Release 4 (DR4, 5.5-year baseline) and Data Release 5 (DR5, 10.5-year baseline).

\section{Semi-analytic model}
\label{sec:analytic_model}

We begin with a semi-analytic model that is simple enough to permit rapid calculations, facilitate understanding, and set expectations for the more realistic models presented in Sections~\ref{sec:astrom_data} and \ref{sec:mock_catalog}.

\subsection{Structure of the calculation}

Consider a star of mass $M_\star$ at distance $r$ with a single planet of mass $m_p$ on a circular orbit of radius $a$. The maximum amplitude of the star's astrometric displacement is
\begin{eqnarray}
\label{eqn:astom_signal}
\alpha &=& \frac{m_p}{(M_\star+m_p)}\,\frac{a}{r}\\
&\approx&
95.4\,\mu{\rm as}\,
\left(\frac{m_p}{M_{\rm J}}\right)
\left(\frac{a}{{\rm AU}}\right)
\left(\!\frac{M_\star}{M_\odot}\!\right)^{\!\!-1}
\left(\frac{r}{10\,{\rm pc}}\right)^{\!\!-1}.
\end{eqnarray}
Our goal is to predict the characteristics of the planets that {\it Gaia} will detect. Three primary ingredients are needed for this calculation:
\begin{enumerate}

\item A model for the {\bf stellar catalog}, including stellar masses, distances, and apparent magnitudes, which are relevant to planet occurrence and {\it Gaia}'s astrometric precision.

\item A model for {\bf planet occurrence} as a function of the stellar and planetary properties.

\item A model for {\it Gaia}'s {\bf detection efficiency} as a function of those same properties.

\end{enumerate}

We aim to calculate the function
\begin{equation}
\Gamma_{M_\star,m_p,a}^{\rm det} \equiv \frac{dN_p^\mathrm{det}(M_\star,m_p,a)}{dM_\star\,dm_p\,da}~,
\end{equation}
which is the expected number of detected planets with mass $m_p$ and orbital radius $a$ around stars of mass $M_\star$, per unit interval of the three input variables $M_\star$, $m_p$, and $a$. We have already simplified the calculation by assuming the orbits are circular. Integrating $\Gamma_{M_\star,m_p,a}^\mathrm{det}$ over the desired range of $M_\star$, $m_p$, and $a$ yields the expected number of detected planets. We can separate the three main ingredients of the calculation by writing this function as a volume integral:
\begin{align}
\label{eqn:volume-integral}
\Gamma_{M_\star, m_p, a}^\mathrm{det} = \int 
&\frac{dN_\star(M_\star, \vec{r})}{dM_\star\,dV}\,
\frac{dN_\mathrm{p}(M_\star, m_p, a, \vec{r})}{dN_\star\,dm_p\,da} \nonumber \\
&\mathrm{P}^\mathrm{det}(M_\star, m_p, a, r) \, dV.
\end{align}
The first factor in the integrand is the volumetric stellar mass function (VMF$_\star$), defined in terms of $dN_\star$, the number of stars per cubic parsec with mass between $M_\star$ and $M_\star\,{+}\,dM_\star$. The second factor is the planet occurrence function. Here, $dN_p$ represents the average number of planets with mass between $m_p$ and $m_p\,{+}\,dm_p$ and orbital radius between $a$ and $a\,{+}\,da$ that exist around a star of mass $M_\star$. The third factor is the probability that a planet of mass $m_p$ and orbital radius $a$ would be detected around a star of mass $M_\star$ located at a distance $r$.

\begin{figure}
\centering
\includegraphics[width=0.475\textwidth]{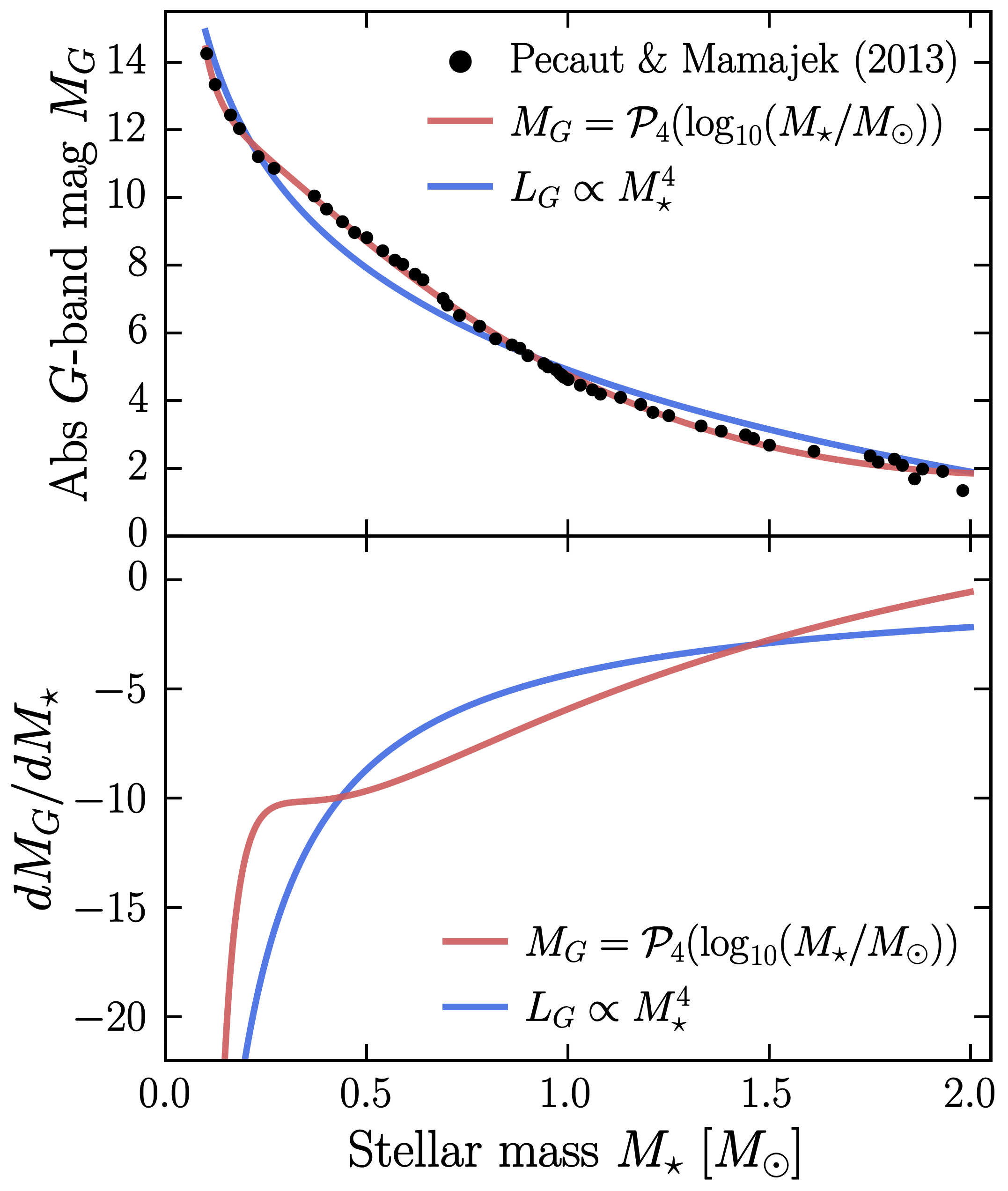}
\caption{The top panel shows the absolute $G$-band magnitude versus stellar mass for main-sequence stars, from \citet{Pecaut&Mamajek2013} (and online updates). The best-fit fourth-order polynomial is plotted in red. The blue curve is the best-fit single-parameter function $L_G \propto M_\star^4$ or, equivalently, $M_G\,{=}\,-10\log_{10}(M_\star/M_\odot)\, {+}\,C$. The bottom panel shows the derivative $d M_G/dM_\star$, in units of $M_\odot^{-1}$, which is needed to convert the luminosity function into a mass function.}
\label{fig:mass-mag}
\end{figure}

\subsection{Stellar properties}

Three stellar properties are relevant to our astrometric planet detection forecasts: mass ($M_\star$), absolute {\it Gaia}-band magnitude ($M_G$), and distance from the Sun ($r$). Both the astrometric signal amplitude and the planet occurrence function depend on $M_\star$. The signal amplitude also depends on $r$, and the achievable astrometric precision depends on both $M_G$ and $r$.

Throughout this paper, our attention will be restricted to main-sequence stars with masses between $0.1$ and $2$~$M_\odot$.\footnote{Widening the mass range would be unlikely to enlarge the expected yield much, because our results and prior studies have shown that planet sensitivity falls off sharply outside this range (see \citealt{Casertano2008, Perryman2014}; Section~\ref{sec:inject_recover_results}).} Several portions of our analysis will require a conversion between $M_\star$ and $M_G$. The top panel of Figure~\ref{fig:mass-mag} shows $M_G$ versus $M_\star$ for zero-age main-sequence stars based on a tabulation originally provided by \citet{Pecaut&Mamajek2013}.\footnote{Taken from \url{https://www.pas.rochester.edu/~emamajek/EEM_dwarf_UBVIJHK_colors_Teff.txt} on June 23rd, 2025.} Also plotted is the best-fit quartic function, which is used in subsequent steps of the calculation. A simpler mass-magnitude relation in which $L_G\,{\propto}\,M_\star^4$ is also shown for reference.

\begin{figure}
\centering
\includegraphics[width=0.475\textwidth]{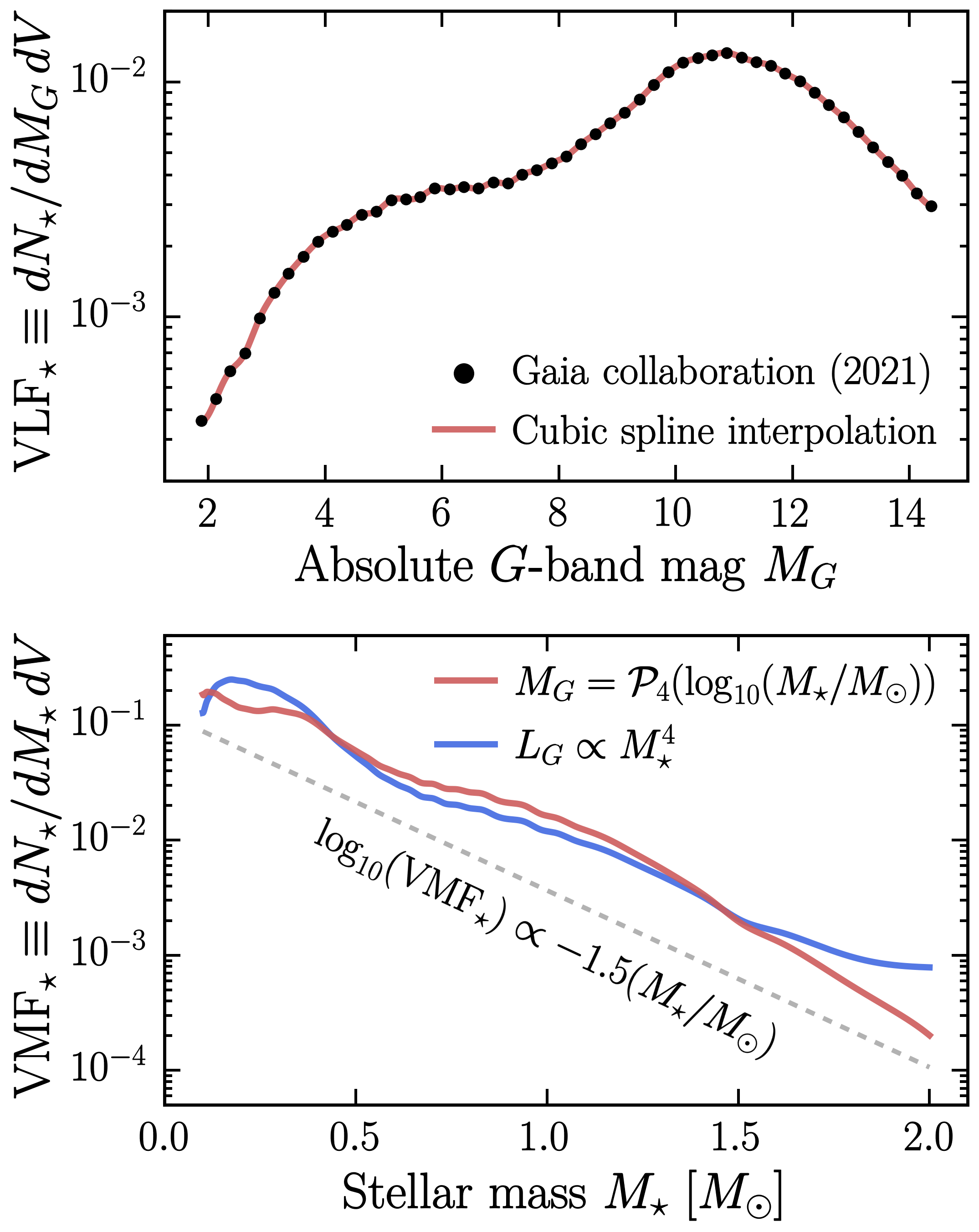}
\caption{The top panel shows the $G$-band volumetric luminosity function for main-sequence stars within $100$~pc \citep{GaiaCollaboration2021}. A cubic spline interpolation is shown in red. The bottom panel shows the corresponding volumetric mass function, based on the two choices for the mass-magnitude relations in Figure~\ref{fig:mass-mag}. For reference, a simple exponential approximation is plotted in gray. In both panels, the units of volume and mass are pc$^3$ and $M_\odot$, respectively.}
\label{fig:VLF_VMF}
\end{figure}

Next, we require a model for the VMF$_\star$. In full generality, the VMF$_\star$ depends on stellar mass, age, composition, and location within the Galaxy. For simplicity, we will not consider age or composition, and we will assume that the mass dependence and spatial dependence are separable:
\begin{equation}
\label{eqn:VMF_sep}
    {\rm VMF}_\star (M_\star,\vec{r}) =
    {\rm VMF}_\star (M_\star) \times f_\star(\vec{r})~.
\end{equation}

To construct a model for the VMF$_\star$, we begin with the precise stellar luminosity function provided by the {\it Gaia} team. The top panel of Figure~\ref{fig:VLF_VMF} shows the volumetric stellar luminosity function (VLF$_\star$) for main-sequence stars within $100$~pc, extracted from Figure~16 of \citet{GaiaCollaboration2021}. The red curve shows a cubic spline fit, which we used to interpolate between the tabulated VLF$_\star$ data. The bottom panel shows the corresponding VMF$_\star$, computed as
\begin{equation}
\mathrm{VMF}_\star(M_\star) = 
\frac{dN_\star[M_G(M_\star)]}{dM_G\,dV}\,\left| \frac{dM_G(M_\star)}{dM_\star}\right|.
\end{equation}
The first term is the VLF$_\star$ and the second term is calculated using the quartic mass-magnitude relation described above (see the bottom panel of Figure~\ref{fig:mass-mag}). The resulting VMF$_\star$ is displayed as a red curve in the bottom panel of Figure~\ref{fig:VLF_VMF}. For reference, the blue curve shows the result of using $L_G\,{\propto}\,M_\star^4$ instead of the quartic $M_\star$-$M_G$ relation, and the dashed line shows a simple exponential approximation for the VMF$_\star$.

\subsection{Detection process}

For this semi-analytic model, we assumed that planet detection is assured if the calculated signal-to-noise ratio {\it per single astrometric data point} (hereafter, \snr) exceeds a threshold value, and is otherwise impossible. The main reason for using \snr\, rather than a statistic based on all the data, was to allow direct comparisons with previous authors \citep[see, e.g.,][]{Lattanzi2000, Casertano2008, Perryman2014}. \snr\ also offers the advantage of simplicity, and we expect there to be a tight correlation between \snr\ and the evidence for the planet based on the data. This expectation is borne out by more detailed models (see \citealt{Perryman2014} and Section~\ref{sec:astrom_data}). The criterion for detection is
\begin{align}
\label{eqn:SNR_criterion}
\mathrm{SNR}_1 &> N_\sigma\,,~\mathrm{where} \\
\label{eqn:SNR_definition}
\mathrm{SNR}_1 &\equiv \frac{\alpha}{\sigma_{\rm fov}(G)} \approx \frac{m_p}{M_\star}\,\frac{a}{r} \frac{1}{\sigma_{\rm fov}(G)}~.
\end{align}
Here, $\sigma_\mathrm{fov}$ is the expected uncertainty in a single angular measurement (i.e., a single field-of-view crossing), and $N_\sigma$ is a constant that sets the detection threshold. Here, we have assumed that $m_p \ll M_\star$. Based on the orbit-fitting experiments in Section~\ref{sec:astrom_data}, we adopted nominal values of $N_\sigma\,{=}\,1.5$ for DR4 and $N_\sigma\,{=}\,1.0$ for DR5. These values are roughly equivalent to requiring $\Delta \chi^2\,{>}\,50$ between models with and without a planet, as will be shown later. In many of the expressions derived below, we include the explicit dependence on $N_\sigma$ to show the effects of adopting a different threshold.

\begin{figure}
\centering
\includegraphics[width=0.475\textwidth]{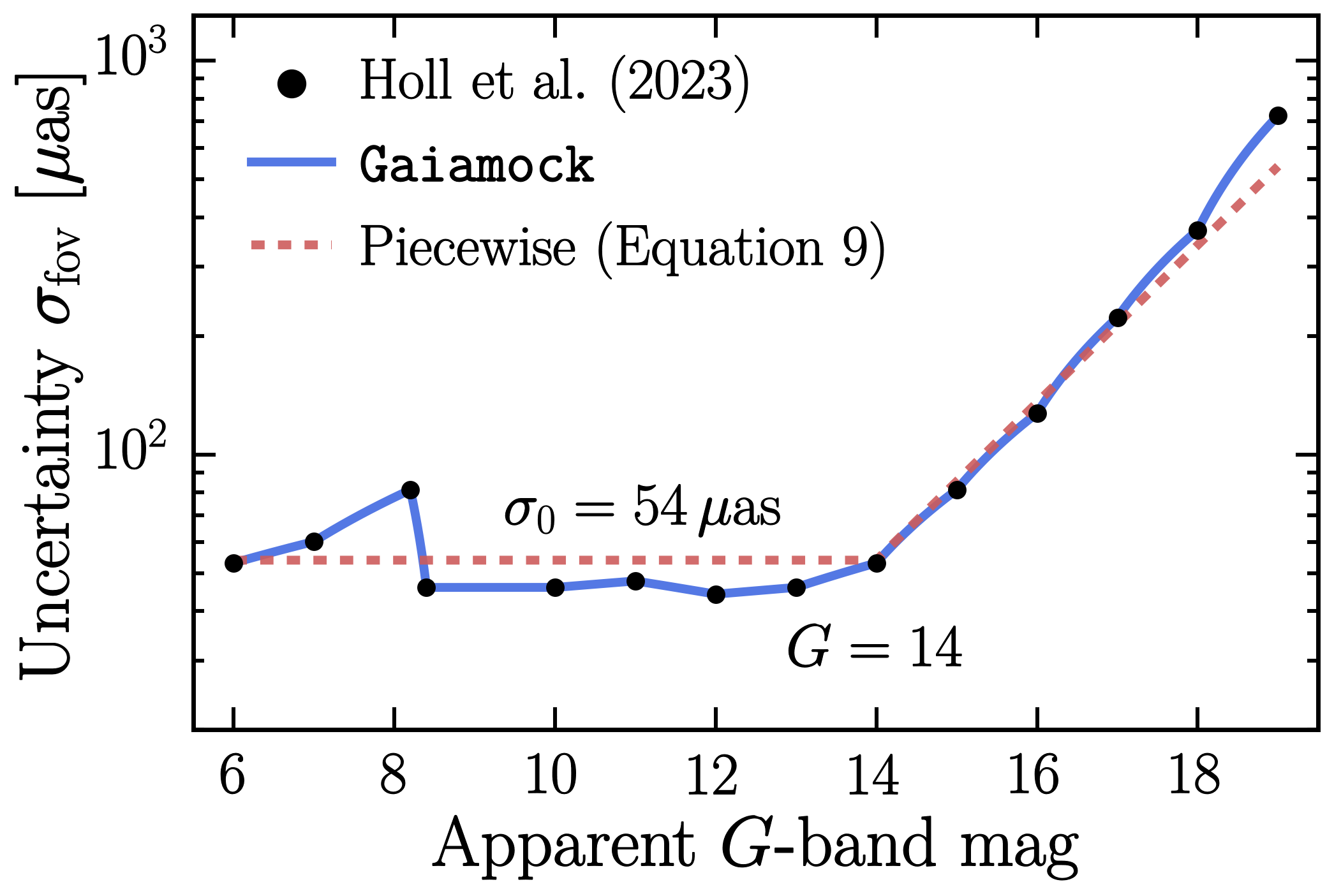}
\caption{Model for the astrometric precision of individual {\it Gaia} measurements (i.e., field-of-view crossings) as a function of apparent $G$-band magnitude. DR3 astrometric uncertainties from \citet{Holl2023a} are shown in black, along with the interpolation used by the \texttt{Gaiamock} code (\citealt{ElBadry2024}; see Section~\ref{sec:astrom_data}). Our simple piecewise model (Equation~\ref{eqn:astrom_unc}; red) was fitted to the \citet{Holl2023a} data. For bright stars ($G\,{<}\,14$), the astrometric uncertainty is approximately constant, whereas for fainter stars ($G\,{>}\,14$), the astrometric uncertainty rises with $G$ in the manner expected due to photon-counting noise.}
\label{fig:astrom_unc}
\end{figure}

{\it Gaia}'s astrometric precision is chiefly a function of the source's apparent $G$-band magnitude ($G$). Figure~\ref{fig:astrom_unc} shows the astrometric precision achieved in DR3 \citep{Holl2023a}, along with the interpolation employed by the \texttt{Gaiamock} code \citep{ElBadry2024}.\footnote{We followed \citet{ElBadry2024} by assuming that each of {\it Gaia}'s eight CCD measurements are independent, and therefore the per-FOV uncertainty is smaller than the per-CCD uncertainty by a factor of $\sqrt{8}$.} We modeled the dependence of {\it Gaia}'s astrometric uncertainty on $G$ as a piecewise function that is flat for stars with $G\,{<}\,14$, and thereafter rises in the manner expected from photon-counting noise:
\begin{equation}
\label{eqn:astrom_unc}
    \sigma_{\rm fov}(G) = \sigma_0 \times \max\left[1,~10^{0.2(G-14)}\right]~.
\end{equation}
The noise floor $\sigma_0\,{=}\,54\,\mu{\rm as}$ and the breakpoint
magnitude $G\,{=}\,14$ were determined by performing a least-squares fit to the data from \citet{Holl2023a} shown in Figure~\ref{fig:astrom_unc}. Over the range $6\,{\le}\,G\,{\le}\,19$, this simple piecewise model for $\sigma_\mathrm{fov}$ has a mean difference of just $12$\% from the interpolation employed by \texttt{Gaiamock}. Note that in the faint-star regime, we can use the standard relation $G\,{=}\,M_G\,{+}\,5\log_\mathrm{10}{(r/10\,{\rm pc})}$ to rewrite the per-point uncertainty as
\begin{equation}
\label{eqn:astrom_unc_faintlimit}
\sigma_{\rm fov} = (54\,\mu{\rm as}) \left(\frac{r}{10~{\rm pc}} \right)~10^{0.2(M_G - 14)},
\end{equation}
which makes it more obvious that the uncertainty scales with distance for a star of a given luminosity. The astrometric precision worsens for stars brighter than $G\,{\lesssim}\,6$ due to calibration-related systematic errors \citep{Holl2023a}. We did not account for this extra source of noise because relatively few planet detections are expected around such bright stars (see Section~\ref{sec:mock_catalog}). Likewise, we do not expect many planet detections around stars fainter than $G\,{\gtrsim}\,19$.

The planet's orbital period is also relevant to the detection probability. Although long-period planets produce large-amplitude astrometric signals (Equation~\ref{eqn:astom_signal}), when the observing baseline does not cover a full orbit, the star's orbital motion and proper motion are difficult to distinguish \citep{Casertano2008, Gould2008, Perryman2014}. Motivated by the orbit-fitting experiments presented in Section~\ref{sec:astrom_data}, we adopted a maximum orbital period of $4.0$~years for DR4 and $9.5$~years for DR5.

\subsection{Limiting distance and magnitude}

Equations~\ref{eqn:SNR_definition} and \ref{eqn:astrom_unc_faintlimit} imply that the \snr\ is a strictly decreasing function of $r$. Therefore, for fixed values of $M_\star$, $m_p$, and $a$, requiring the \snr\ to exceed a threshold value is equivalent to requiring the star to be sufficiently nearby: $r\,{<}\,r_\mathrm{max}(M_\star,m_p,a)$. There are two cases:
\begin{enumerate}

\item For bright stars ($G\,{<}\,14$), the astrometric uncertainty is independent of distance. Using Equation~\ref{eqn:SNR_definition} with $\sigma_{\rm fov} = \sigma_0$, the detection criterion is
\begin{equation}
\label{eqn:snr-threshold-case-1}
    \left( \frac{m_p}{M_\star}\right)
    \left(\frac{a}{r}\right)
    \left(\frac{1}{\sigma_0} \right) > N_\sigma~,
\end{equation}
which can be solved for $r$ to find
\begin{eqnarray}
\label{eqn:rmax-case-1}
r &<& 
\frac{1}{N_\sigma}
\left(\frac{m_p}{M_\star}\right)
\left(\frac{a}{\sigma_0}\right)\\
\label{eqn:rmax-scaling-case-1}
&<&52\,{\rm pc}
\left(\frac{m_p}{M_{\rm J}}\right)
\!\left(\frac{a}{1\,{\rm AU}}\right)
\!\left(\frac{\!0.34\,M_\odot\!\!}{M_\star}\right)~,
\end{eqnarray}
where in the last equation we used $N_\sigma\,{=}\,1.0$ and $\sigma_0\,{=}\,54\,\mu$as. We chose a nominal value of $0.34~M_\odot$ because such a star has $G\,{\approx}\,14$ at the nominal distance of $52$~pc.

\item For faint stars ($14\,{<}\,G\,{<}\,19$), the astrometric uncertainty is proportional to distance. Using Equations~\ref{eqn:SNR_definition} and~\ref{eqn:astrom_unc_faintlimit}, the detection criterion is
\begin{equation}
\label{eqn:snr-threshold-case-2}
    \left( \frac{m_p}{M_\star}\right)
    \!\!\left(\!\frac{a \cdot 10\,{\rm pc}}{r^2}\!\right)
    \!\!\left(\!\frac{1}{\sigma_0}\!\right)
10^{-0.2(M_G-14)} > N_\sigma,
\end{equation}
which can be solved for $r$ to find
\begin{equation}
\label{eqn:rmax-case-2}
r < \sqrt{ \frac{m_p}{M_\star} \frac{a \cdot 10\,{\rm pc}} {N_\sigma\,\sigma_0 }} ~10^{-0.1(M_G-14)},
\end{equation}
or, as a scaling relation,
\begin{equation}
\label{eqn:rmax-scaling-case-2}
r < 52\,{\rm pc}\, \sqrt{ \frac{m_p}{M_{\rm J}} \, \frac{a}{1\,{\rm AU}} \, \frac{0.34\,M_\odot}{M_\star}}~10^{-0.1(M_G-10.4)},
\end{equation}
where again we used $N_\sigma\,{=}\,1.0$ and $\sigma_0\,{=}\,54\,\mu$as. Note that the last factor in this equation is proportional to $L_G^{1/4}$, where $L_G$ is the {\it Gaia}-band luminosity, which will be useful below.

\end{enumerate}

\begin{figure*}
\centering
\includegraphics[width=0.95\textwidth]{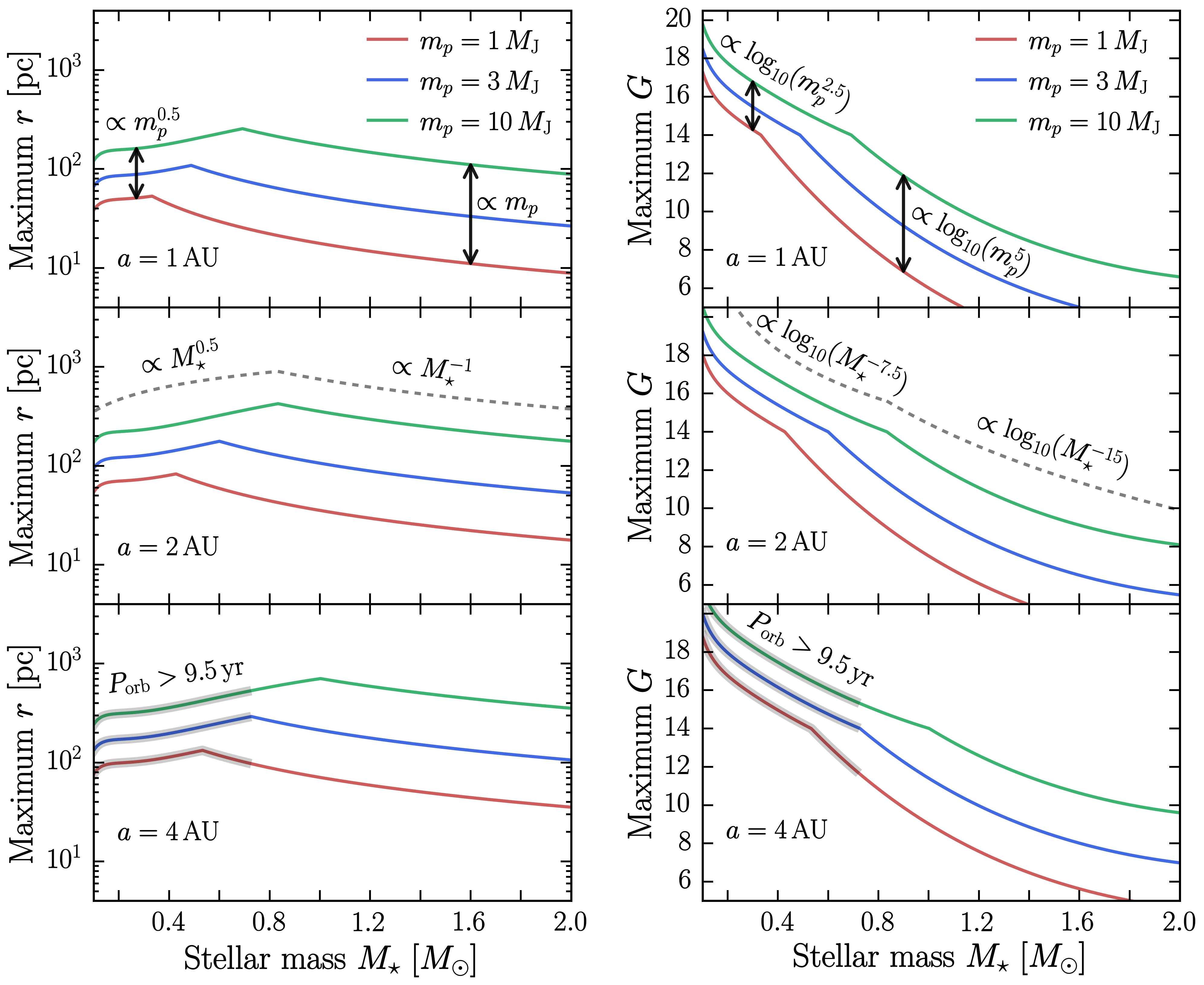}
\caption{Predictions for the maximum distance (left column) and maximum apparent $G$-band magnitude (right column) of stars for which {\it Gaia} can detect planets in DR5. Within each panel, curves are shown for three different planet masses. Each of the three panels is for a different orbital distance. The kinks are due to the break in our model for {\it Gaia}'s astrometric precision as a function of apparent magnitude (Figure~\ref{fig:astrom_unc}). In the middle panel of each column, the gray dashed curve shows the scaling expected from a simple analytic model. In the top panels, the arrows indicate how the vertical spacing between curves scales with planet mass. In the bottom panels, the gray bands highlight the cases for which the orbital period exceeds $9.5$~years, which will make astrometric detection challenging. The most distant stars for which {\it Gaia} can detect planets are several hundred parsecs away. The faintest $G$ magnitude depends sensitively on stellar mass.}
\label{fig:rmax_Gmax}
\end{figure*}

We are now prepared to calculate {\it Gaia}'s limiting distance and limiting apparent magnitude for detecting planets, as a function of $M_\star$, $m_p$, and $a$. The left column of Figure~\ref{fig:rmax_Gmax} shows the limiting distance versus stellar mass, for different choices of planet mass and orbital distance. The shapes of these curves can be understood analytically. The kinks are caused by the kink in our adopted noise model (Figure~\ref{fig:astrom_unc}). High-mass stars are in the bright-star regime, for which $r_\mathrm{max} \propto\,M_\star^{-1} m_p\,a$ (Equation~\ref{eqn:rmax-scaling-case-1}). The scaling with $M_\star^{-1}$ can be observed at high stellar masses (see the dashed gray line in the middle plot). The linear scaling with $m_p$ is evident from the vertical spacings between differently colored curves. The linear scaling with $a$ can be confirmed by comparing curves of the same color across different panels.

Low-mass stars are in the faint-star regime, which is more complicated because the limiting distance depends on both stellar mass and luminosity. However, using the approximation $L_G\,{\propto}\,M_\star^4$, Equation~\ref{eqn:rmax-scaling-case-2} implies $r_\mathrm{max} \propto\,M_\star^{0.5}\,m_p^{0.5}\,a^{0.5}$. The left column of Figure~\ref{fig:rmax_Gmax} reveals that the scaling with $M_\star^{0.5}$ is roughly accurate (again, refer to the dashed gray line in the middle panel). The square-root scaling with $m_p$ is the reason why the differently colored curves become more closely spaced for low stellar masses. The square-root scaling with $a$ can also be verified by comparing different panels.

Overall, the left column of Figure~\ref{fig:rmax_Gmax} shows that Jupiter-mass planets can only be detected around stars within $100$~pc, regardless of stellar mass and orbital distance. However, {\it Gaia} will be sensitive to super-Jupiters ($m_p\,{\approx}\,10M_\mathrm{J}$) at several AU for stars within ${\sim}\,300$~pc, across a wide range of stellar masses.

The right column of Figure~\ref{fig:rmax_Gmax} shows the limiting apparent $G$ magnitude versus stellar mass, for different planet masses and orbital distances. These plots highlight the strong dependence of the limiting magnitude on stellar mass, contrasting with the fairly weak dependence of the limiting distance on stellar mass. The scalings can be understood analytically by recalling the relationship between $G$ and $r$. For $G\,{<}\,14$, $G_\mathrm{max}$ scales as ${\propto}\,\log_{10}(M_\star^{-15}\,m_p^5\,a^5)$. For $G\,{>}\,14$, $G_\mathrm{max}$ scales as ${\propto}\,\log_{10}(M_\star^{-7.5}\,m_p^{2.5}\,a^{2.5})$. The highest-mass stars ($2$~$M_\odot$) can only be searched for planets when $G\,{\lesssim}\,10$, whereas the lowest-mass stars ($0.2$~$M_\odot$) can be searched for planets even when they are as faint as $G\,{\approx}\,18$.

\subsection{Integrating over volume}

Next, we combined our model for {\it Gaia}'s sensitivity to planets with knowledge of the local stellar population to predict the number of stars {\it Gaia} can effectively search for planets. Assuming that the planet occurrence function is independent of location, we can pull it out of the volume integral in Equation~\ref{eqn:volume-integral}, yielding
\begin{align}
\label{eqn:volume-integral-2}
\Gamma_{M_\star,m_p,a}^{\rm det} = 
&\frac{dN_{\rm p}(M_\star,m_p,a)}{dN_\star\,dm_p\,da}\, \nonumber \\
&\int 
{\rm VMF}_\star(M_\star,\vec{r})~
{\rm P}^{\rm det}(M_\star,m_p,a,r) \, dV~.
\end{align}
The integral gives the number of stars that {\it Gaia} can search for planets of mass $m_p$ and orbital radius $a$, per unit stellar mass. We will refer to this integral as the mass function of {\it searchable} stars:
\begin{align}
{\rm MF}_\star^{\rm s}(M_\star,m_p,a) \equiv& 
\int 
{\rm VMF}_\star(M_\star,\vec{r})\,
{\rm P}^{\rm det}(M_\star,m_p,a,\,r) \, dV \nonumber \\
=& \frac{dN^{\rm s}_\star(M_\star,m_p,a)}{dM_\star}.
\end{align}
Integrating over stellar mass yields the total number of searchable stars.

Since we are assuming $\mathrm{P}^{\rm det}$ is unity when $r\,{<}\,r_{\rm max}$ and zero otherwise, we can rewrite $\mathrm{MF}^{\rm s}_\star(M_\star,m_p,a)$ as
\begin{align}
\label{eqn:searchable_V}
{\rm MF}_\star^{\rm s} =&
\int \!d\Omega
\int_0^{r_{\rm max}(M_\star,m_p,a)}
\!\!\!\!\!\!{\rm VMF}_\star(M_\star,\vec{r})\, r^2 \,dr \nonumber \\
=&
{\rm VMF}_\star(M_\star)
\int \!d\Omega
\int_0^{r_{\rm max}(M_\star,m_p,a)}
\!\!\!\!\!\!\, f_\star(\vec{r})\,r^2 \,dr \nonumber \\
\equiv&
{\rm VMF}_\star(M_\star) \times V^{\rm s}~,
\end{align}
where, in the second line, we assumed the VMF$_\star$ to be independent of location (Equation~\ref{eqn:VMF_sep}). In the third line, we implicitly defined the {\it searchable volume} $V^{\rm s}$ for a given choice of $\{M_\star,m_p,a\}$. Because $r_\mathrm{max}$ extends to several hundred parsecs in some cases (Figure~\ref{fig:rmax_Gmax}), the vertical structure of the Galaxy is relevant. Assuming for simplicity that the volumetric stellar mass function decreases exponentially with height $z$ above the disk midplane, the searchable volume becomes
\begin{align}
\label{eqn:searchable_V_exponential}
V^\mathrm{s} =& \int e^{-|z|/H} dV \nonumber \\
  =& \int_0^{\pi} \!\!2\pi \sin\theta\,d\theta \int_0^{r_{\rm max}(M_\star,m_p,a)}
  \!\!\!e^{-|r\cos\theta|/H}
  \,r^2 dr \nonumber \\
  =& \frac{4\pi}{3} r_{\rm max}^3 \left[ 3~\frac{\frac{\xi^2}{2} +
    e^{-\xi}(1+\xi) - 1}{\xi^3} \right]~.
\end{align}
Here, $H$ is the vertical scale height of the Galactic disk, and we have defined $\xi\,{\equiv}\,r_\mathrm{max}/H$. Notice that in the nearby limit ($\xi\,{\rightarrow}\,0$), $V^\mathrm{s}$ is simply the volume of a sphere of radius $r_\mathrm{max}$. In the faraway limit ($\xi\,{\rightarrow}\,\infty$), $V^\mathrm{s}$ approaches the volume of a cylinder with radius $r_\mathrm{max}$ and height $2H$. We adopted a scale height of $H\,{=}\,300$\,pc. The predictions of the model depend somewhat on this choice; ${\rm MF}_\star^{\rm s}(M_\star)$ differs by up to $30$\% when $H$ is changed to $200$~pc.

\begin{figure*}
\centering
\includegraphics[width=0.95\textwidth]{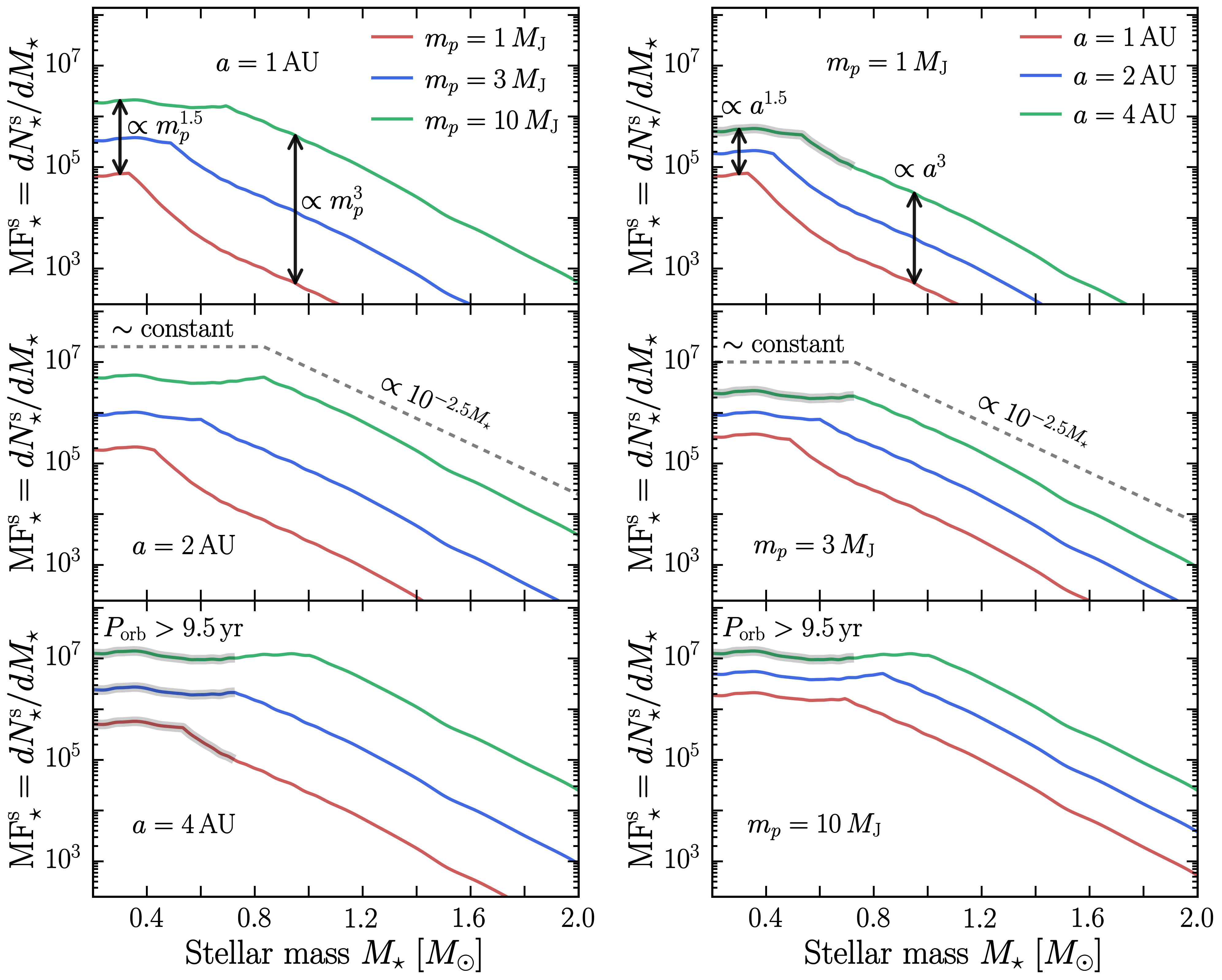}
\caption{Predictions for the mass function of stars that could host a {\it Gaia}-detectable planet (MF$_{\star}^{\rm s}$) in DR5. Plots in the left column show MF$_{\star}^{\rm s}$, in units of $M_\odot^{-1}$, versus stellar mass for a fixed orbital radius and three different choices of planet mass. Plots on the right show MF$_\star$ versus stellar mass for planets with a fixed mass and three different choices of orbital radius. As the stellar mass rises, MF$_\star$ is initially constant before suddenly falling as $10^{-2.5M_\star}$. The transition occurs when the searchable stars cross $G\,{=}\,14$, the breakpoint in our model of {\it Gaia}'s astrometric precision (see Figure~\ref{fig:astrom_unc}). The black arrows illustrate how MF$_{\star}^{\rm s}$ scales with $m_p$ and $a$. Millions of stars could host detectable planets, most of which are less massive than the Sun.}
\label{fig:num_searchable}
\end{figure*}

Figure~\ref{fig:num_searchable} shows the mass function of searchable stars as a function of $M_\star$, $m_p$, and $a$, computed using Equations~\ref{eqn:searchable_V} and~\ref{eqn:searchable_V_exponential}, where $r_\mathrm{max}$ comes from Equations~\ref{eqn:rmax-scaling-case-1} and~\ref{eqn:rmax-scaling-case-2}, and the VMF$_\star$ is the red curve in Figure~\ref{fig:VLF_VMF}. As the stellar mass is increased, the mass function of searchable stars is initially constant. After reaching a critical stellar mass, the mass function of searchable stars begins declining exponentially. The critical stellar mass separating these two regimes corresponds to the kink in $r_\mathrm{max}$ as a function of stellar mass (Figure~\ref{fig:rmax_Gmax}), which in turn corresponds to the $G\,{=}\,14$ breakpoint of the noise model (Figure~\ref{fig:astrom_unc}). Figure~\ref{fig:num_searchable} also shows that the number of searchable stars rises sharply with planet mass and orbital distance. All of these scalings can be understood analytically, as shown below.

For stars with $G\,{<}\,14$, there are three relevant relations: $\mathrm{MF}_\star^{\rm s}$ is approximately proportional to $\mathrm{VMF}_\star\,r_\mathrm{max}^3$ (Equations~\ref{eqn:searchable_V} and \ref{eqn:searchable_V_exponential}), $\mathrm{VMF}_\star$ scales approximately as $10^{-1.5 M_\star}$ (Figure~\ref{fig:VLF_VMF}), and $r_\mathrm{max}$ scales as $M_\star^{-1}\,m_p\,a$ (Equation~\ref{eqn:rmax-scaling-case-1}). Combining these relations, we obtain
\begin{align}
\label{eqn:mass_func_bright}
\mathrm{MF}_\star^{\rm s} &\propto 10^{-1.5 M_\star} \left(\frac{m_p a}{M_\star}\right)^{\!3}\nonumber \\
&\propto 10^{-2.5M_\star}~m_p^3~a^3 ~{\rm (bright~limit;}~G < 14{\rm )}~,
\end{align}
where in the second line we approximated $(M_\star/M_\odot)^{-3}\,{\approx}\,9\,{\times}\,10^{-M_\star/M_\odot}$, which is valid to first order when $M_\star\,{\approx}\,1.3\,M_\odot$. These expectations are borne out in Figure~\ref{fig:num_searchable}.

For stars with $G\,{>}\,14$, the scalings $\mathrm{MF}_\star^{\rm s}\,{\propto}\,\mathrm{VMF}_\star\,r_\mathrm{max}^3$ and $\mathrm{VMF}_\star(M_\star)\,{\propto}\,10^{-1.5 M_\star}$ are still valid, but in this case $r_\mathrm{max}$ scales as $M_\star^{0.5}\,m_p^{0.5}\,a^{0.5}$ (Equation~\ref{eqn:rmax-scaling-case-2}). It follows that
\begin{align}
\label{eqn:mass_func_faint}
\mathrm{MF}_\star^{\rm s} &\propto 10^{-1.5 M_\star} \left(m_p\, a\, M_\star\right)^{1.5} \nonumber \\
&\propto m_p^{1.5} a^{1.5}~{\rm (faint~limit;~ G > 14)~,}
\end{align}
where we have used the approximation $10^{-1.5(M_\star/M_\odot)}\,{\approx}\,16\,(M_\star/M_\odot)^{-1.5}$, valid to first order when $M_\star\,{\approx}\,0.43\,M_\odot$. In this approximation, the stellar mass dependence drops out. Thus, the plateau in MF$_\star$ at low stellar mass is caused by a coincidental cancellation between the stellar-mass dependence of $r_\mathrm{max}$ and that of the VMF$_\star$.

\subsection{Planet occurrence model}
\label{sec:planet_occ}

Forecasting planet detections requires knowledge of both the number of searchable stars and the planet occurrence function. Around FGKM-type stars, the occurrence statistics of planets with the relevant masses (${\sim}\,1$\,--\,$13\,M_\mathrm{J}$) and orbital distances (${\sim}\,0.5$\,--\,$5$\,AU) have been measured by long-term radial-velocity (RV) surveys \citep[e.g.,][]{Cumming2008, Reffert2015, Wittenmyer2016}. For our forecasts, we adopted the occurrence function from the California Legacy Survey (CLS; \citealt{Rosenthal2021, Fulton2021}). The CLS reported an occurrence rate of $14\,{\pm}\,2$ giant planets with orbital separations $2$\,--\,$8$\,AU per $100$ FGKM stars, based on an RV survey of $719$ stars. They modeled the occurrence rate density of planets with masses between $30$ and $6000~M_\oplus$ and semi-major axes between $0.1$ and $30$~AU using a broken power-law function,
\begin{equation}
\label{eqn:F21_occ_rate}
\frac{dN_p}{dN_\star\,d \ln a\, d \ln m_p} = 
C\,
\left(\frac{a}{\mathrm{AU}}\right)^\beta \left[1-e^{-(a/a_0)^\gamma}\right]~,
\end{equation}
with parameters $C\,{=}\,0.15^{+0.24}_{-0.09}$, $\beta\,{=}\,-0.86^{+0.41}_{-0.41}$, $a_0\,{=}\,3.6^{+2.0}_{-1.8}$\,AU, and $\gamma\,{=}\,1.59^{+0.36}_{-0.33}$.\footnote{\citet{Fulton2021} reported a normalization of $C\,{=}\,350^{+580}_{-220}$, which was specific to their sample size of $719$ stars, the mass range ($30$\,--\,$6000\,M_\oplus$), and the bin width of $\Delta \ln (a)\,{=}\,0.63$. To re-normalize the function as needed here, we divided by $719\,{\times}\,\ln(6000/30)\,{\times}\,0.63\,{\approx}\,2400$.} The uncertainties in the parameters are strongly covariant (see Figure~4 of \citealt{Fulton2021}).

The CLS occurrence rate model does not depend on stellar mass and metallicity, even though giant-planet occurrence is known to vary with those two stellar properties (see \citealt{Johnson2010}). Assuming that the CLS stars and the relevant {\it Gaia} stars have similar metallicity distributions, the CLS model can be used to predict the total number of {\it Gaia} planets without accounting explicitly for metallicity.
This assumption seems reasonable because the metallicity distribution of the CLS stars is centered at $0.0$ with a spread from $-0.5$ to $0.5$ \citep[see Figure~4 of][]{Rosenthal2021}, as observed in the solar neighborhood. However, we must contend with the dependence on stellar mass, because the detectability of planets depends sensitively on stellar mass (see Figures~\ref{fig:rmax_Gmax} and \ref{fig:num_searchable}). The occurrence of giant planets as a function of stellar mass has been investigated extensively, at least as far back as \citet{Laws2003}. The occurrence rate of giant planets with orbital separations less than a few AU has been found to grow approximately linearly with stellar mass up to ${\sim}\,2\,M_\odot$ \citep{Johnson2010, Reffert2015, Jones2016, Ghezzi2018, Wolthoff2022}. To incorporate this trend into our planet occurrence model, we allowed the normalization constant $C$ to depend on stellar mass:
\begin{equation}
\label{eqn:stellar_mass_dep}
\tilde{C}(M_\star) = C \left(\frac{M_\star}{0.9 M_\odot}\right)~.
\end{equation}
The value of $0.9$ was chosen so that our occurrence rate function approximately reproduces the statistics reported by \citet{Fulton2021}. Equations~\ref{eqn:F21_occ_rate} and ~\ref{eqn:stellar_mass_dep} predict $15$ planets per hundred stars with $a\,{=}\,2$\,--$8$\,AU and $m_p\,{=}\,0.3$\,--$13\,M_\mathrm{J}$ over the CLS stellar mass distribution. This function also produces realistic occurrence rates for low-mass stars. Over the range $a\,{=}\,0.1$\,--\,$20$\,AU, $m_p\,{=}\,1.0$\,--\,$13\,M_\mathrm{J}$, and $M_\star\,{=}\,0.1$\,--\,$0.6\,M_\odot$, Equations~\ref{eqn:F21_occ_rate} and ~\ref{eqn:stellar_mass_dep} predict an occurrence rate of $7.7$\%, in agreement with the $6.5\,{\pm}\,3$\% reported by \citet{Montet2014} based on a combination of RV and direct imaging data. See Section~\ref{sec:unc_planet_occ} for discussion on the uncertainty in the planet occurrence function.

\begin{figure*}
\centering
\includegraphics[width=0.95\textwidth]{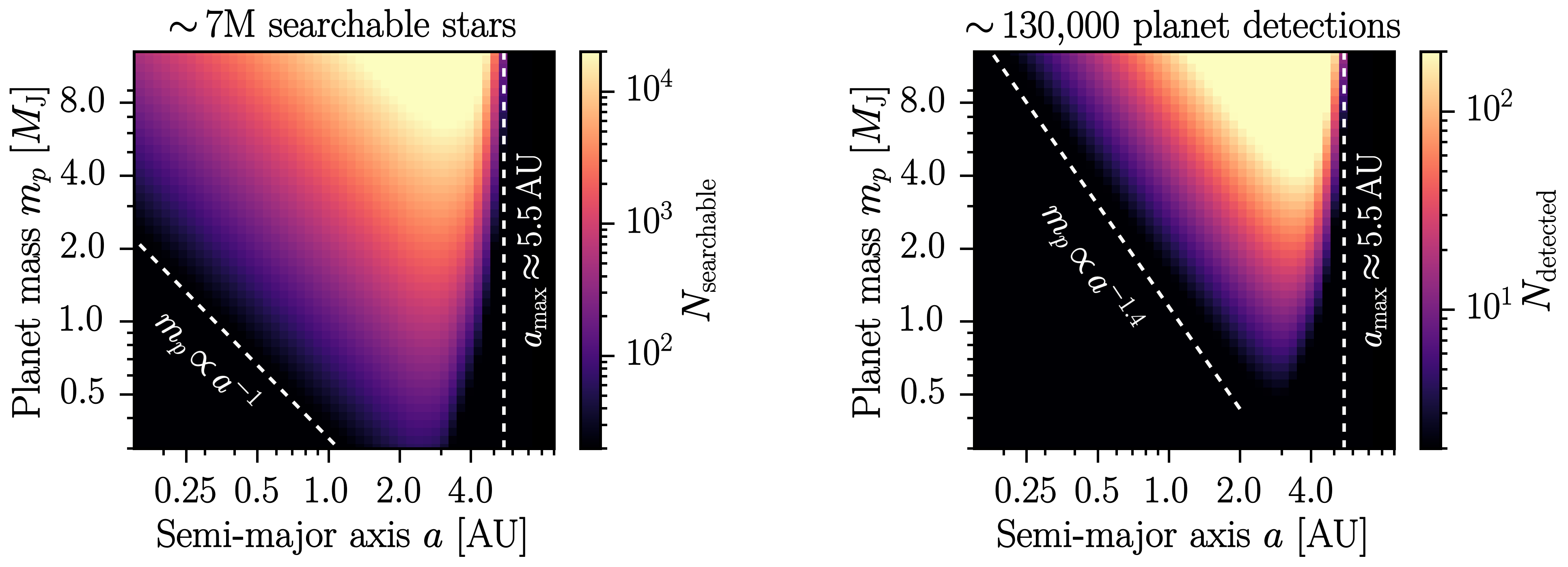}
\caption{Predictions of our semi-analytic {\it Gaia} forecast model for DR5. The left panel shows the number of stars {\it Gaia} will search that could host a detectable planet, as a function of planet mass and semi-major axis. The right panel shows the number of forecasted planet detections as a function of planet mass and semi-major axis. Both adopt a detection threshold of $N_\sigma\,{=}\,1.0$ (see Equation~\ref{eqn:SNR_criterion}). Our model predicts that ${\gtrsim}\,10^6$ stars could host a detectable planet (${\sim}\,1$~million in DR4 and ${\sim}\,7$~million in DR5), and ${\gtrsim}\,10^4$ planets will be detected (${\sim}\,14{,}000$ in DR4 and ${\sim}\,130{,}000$ in DR5). The leftmost ridge comes from the statistics of searchable stars and the planet occurrence function (see Section~\ref{sec:planet_occ}). The fall-off beyond ${\sim}\,5.5$\,AU is due to the restriction $P_\mathrm{orb}\,{\lesssim}\,9.5$\,years so that orbital motion can be distinguished from the star's proper motion (see Section~\ref{sec:max_Porb}).}
\label{fig:detection_grid}
\end{figure*}

\subsection{Planet detections}
\label{sec:analytical_planet_dets}

We are now equipped to provide forecasts for {\it Gaia}'s astrometric planet survey. To begin, we constructed a 3D grid of values of stellar mass, orbital radius, and planet mass, spanning $0.1$\,--$2\,M_\odot$, $0.1$\,--\,$7$\,AU, and $0.3$\,--\,$13\,M_\mathrm{J}$, respectively. At each grid point, we calculated the number of searchable stars by integrating the mass function of searchable stars (Equation~\ref{eqn:searchable_V}) over the relevant range of $M_\star$, $a$, and $m_p$. The left panel of Figure~\ref{fig:detection_grid} shows the grid of searchable stars summed over stellar mass. To compute the expected number of planet detections associated with each grid point, we multiplied the number of searchable stars by the corresponding value of the planet occurrence function. The right panel of Figure~\ref{fig:detection_grid} shows the results as a function of orbital radius and planet mass, summed over stellar mass. For these calculations, we also required $P_\mathrm{orb}\,{<}\,9.5$\,years, the approximate maximum orbital period for planets detectable with {\it Gaia} DR5 data.

In total, our semi-analytic model predicts that {\it Gaia} DR5 will include data for ${\sim}\,7$~million stars around which a planet could be detected, and will result in ${\sim}\,130{,}000$ planet detections. As illustrated in Figure~\ref{fig:detection_grid}, the vast majority of the detected planets will be more massive than Jupiter and will have orbital distances of ${\sim}\,2$\,--\,$5$\,AU. Most of the detected planets will belong to stars with masses near or just below the Sun's mass ($M_\star\,{\approx}\,0.9\,M_\odot$); only $20$\% of detections occur around M-dwarfs ($M_\star\,{<}\,0.6\,M_\odot$). For a given type of planet, the astrometric signal is largest when the host star has the lowest possible mass. Nevertheless, the M-dwarfs do not dominate the predicted catalog of stars with detected planets because: (1) low-mass stars are faint, increasing the astrometric noise (Figure~\ref{fig:astrom_unc}), and (2) giant planets are relatively rare around low-mass stars.

In Figure~\ref{fig:detection_grid}, the searchable stars and the detected planets reside in triangular regions. The boundaries of these regions can be understood analytically. In both plots, the sharp cut-off at large orbital distances is due to the imposition of a maximum orbital period of $9.5$ years, which corresponds to $a\,{\approx}\,5.5$\,AU for the most massive stars that can be effectively searched ($M_\star\,{\approx}\,1.8\,M_\odot$). In the plot of searchable stars (left panel), the slanted boundary obeys $m_p\,{\propto}\,a^{-1}$ because the number of searchable stars depends on $m_p$ and $a$ only through their product $m_p\,a$ (see Equations~\ref{eqn:mass_func_bright} and~\ref{eqn:mass_func_faint}). In the plot of detected planets (right panel), the slanted boundary has a steeper slope because the planet occurrence rate increases with $a$ until $a\,{\approx}\,3$~AU. Expanding the exponential in Equation~\ref{eqn:F21_occ_rate} about $a\,{=}\,0$ yields $C (a/\mathrm{AU})^\beta (a/a_0)^\gamma\,{\propto}\,a^{0.7}$. Including this additional factor yields $N_\mathrm{detected}\,{\propto}\,m^3 a^{3.7}$ for faint stars and $N_\mathrm{detected}\,{\propto}\,m^{1.5} a^{2.2}$ for bright stars. At fixed $N_\mathrm{detected}$, we expect $m_p\,{\propto}\,a^{-1.2}$ and $m_p\,{\propto}\,a^{-1.5}$ in the faint and bright limits, respectively. The intermediate power law $m_p\,{\propto}\,a^{-1.4}$ successfully describes the contours in the planet detection grid.

We repeated these calculations after reducing the maximum allowed orbital period from $9.5$~years to $4.0$~years and increasing the detection threshold from $N_\sigma\,{=}\,1.0$ to $N_\sigma\,{=}\,1.5$, as appropriate for DR4. In this case, the model predicts there will be ${\sim}\,1$\,million searchable stars and ${\sim}\,14{,}000$ planet detections. As before, most of the detected planets will be super-Jupiters around stars that are slightly less massive than the Sun, with somewhat smaller orbital distances ($a\,{=}\,1$\,--\,$3$\,AU). 

The Python code to reproduce the predictions of our semi-analytical model is available on GitHub.\footnote{\url{https://github.com/CalebLammers/GaiaForecasts}}

\subsection{Varying the detection threshold}
\label{sec:analytical_planet_dets_SNR}

\begin{figure}
\centering
\includegraphics[width=0.475\textwidth]{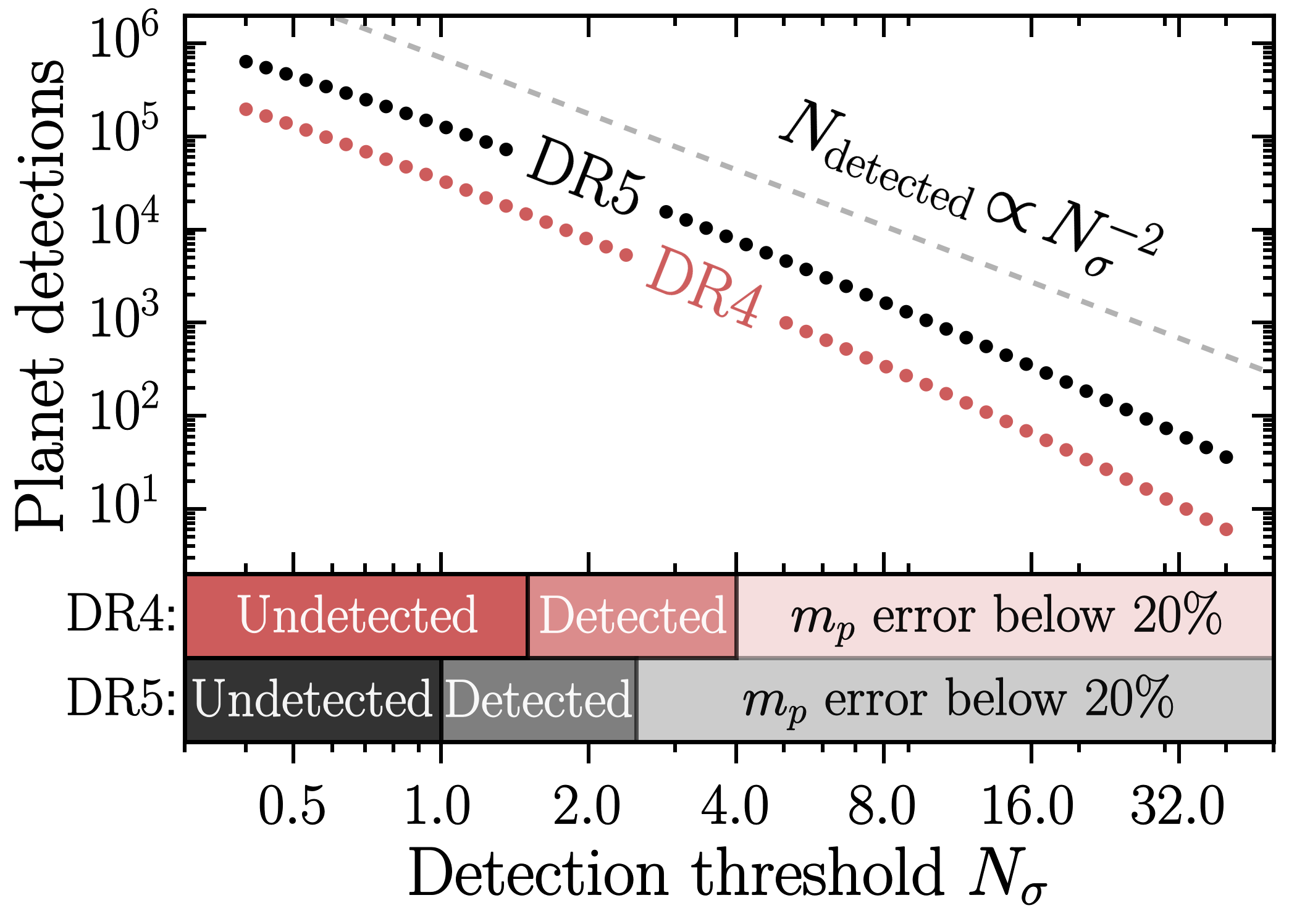}
\caption{The number of planet detections predicted by our semi-analytic model as a function of the detection threshold, $N_\sigma$, defined as the astrometric signal amplitude (Equation~\ref{eqn:astom_signal}) divided by $\sigma_{\rm fov}$, the per-point measurement uncertainty (Equation~\ref{eqn:SNR_criterion}). The number of detections scales approximately as $N_\sigma^{-2}$. The more statistically rigorous criterion $\Delta \chi^2\,{>}\,50$ corresponds to $N_\sigma\,{\gtrsim}\,1.5$ for DR4 and $N_\sigma\,{\gtrsim}\,1.0$ for DR5. To measure planet masses with $20$\% precision for $90$\% of systems requires $N_\sigma\,{\gtrsim}\,4$ for DR4 and $N_\sigma\,{\gtrsim}\,2.5$ for DR5 (see Section~\ref{sec:SNR_chi2_delta}).}
\label{fig:det_vs_Nsigma}
\end{figure}

The predictions shown in Figure~\ref{fig:detection_grid} and earlier figures are based on an assumed detection threshold of $N_\sigma\,{=}\,1.0$ (Equation~\ref{eqn:SNR_criterion}). The simplicity of the semi-analytic model makes it straightforward to study the dependence of the results on $N_\sigma$. Figure~\ref{fig:det_vs_Nsigma} shows the number of detected planets ($N_\mathrm{detected}$) for $50$ values of $N_\sigma$ between $0.4$ and $40$, as predicted using the process described in Section~\ref{sec:analytical_planet_dets}. The red and black curves show the results for DR4 and DR5, for which the restrictions $P_\mathrm{orb}\,{<}\,4.0$\,years and $P_\mathrm{orb}\,{<}\,9.5$\,years were imposed, respectively. The horizontal bars beneath the plot mark the approximate thresholds for secure planet detection and reliable planet mass measurements (see Section~\ref{sec:astrom_data}).

The number of detected planets depends sensitively on $N_\sigma$. For DR5, about $440{,}000$ planets are predicted to satisfy the threshold $N_\sigma\,{>}\,0.5$, whereas only ${\sim}\,200$ planets are expected to have $N_\sigma\,{>}\,20$. We found $N_\mathrm{detected}$ to scale roughly with $N_\sigma^{-2}$, which can be understood in approximate analytic terms. In the bright-star limit ($G\,{<}\,14$), we showed that $r_\mathrm{max} \propto N_\sigma^{-1}$ (Equation~\ref{eqn:rmax-scaling-case-1}), and therefore $N_\mathrm{searchable} \propto r_\mathrm{max}^3 \propto N_\sigma^{-3}$. On the other hand, in the faint-star limit ($G\,{>}\,14$), we showed that $r_\mathrm{max} \propto N_\sigma^{-1/2}$ (Equation~\ref{eqn:rmax-scaling-case-2}) and therefore $N_\mathrm{searchable} \propto N_\sigma^{-3/2}$. Because planet detections occur around a comparable number of stars with $G\,{<}\,14$ and $G\,{>}\,14$ (see Section~\ref{sec:mock_catalog}), the trend turns out to be well described by the single power-law $N_\mathrm{detected}\,{\propto}\,N_\sigma^{-2}$.

\subsection{Uncertainty in the planet occurrence function}
\label{sec:unc_planet_occ}

Whereas the local stellar population and {\it Gaia}'s DR3 astrometric precision are fairly well constrained, the occurrence rate of super-Jupiters on wide orbits is more uncertain. There is uncertainty both in the parameters of the broken power-law function of \citet{Fulton2021} and in the stellar mass dependence introduced via Equation~\ref{eqn:F21_occ_rate}. To quantify the uncertainty in our predictions due to the planet occurrence function, we repeated the calculations described in Section~\ref{sec:analytical_planet_dets} using modified planet occurrence functions. We drew the values of the parameters $C$, $\beta$, $a_0$, and $\gamma$ from the samples of the posterior probability density provided by \citet{Fulton2021}.\footnote{Available at \url{https://github.com/leerosenthalj/CLSII}} We also considered a more general family of stellar mass extrapolations,
\begin{equation}
\label{eqn:stellar_mass_dep_v2}
\tilde{C}(M_\star) = C \left(\frac{M_\star}{K M_\odot}\right)^{\!\eta}~\,
\end{equation}
where $K$ and $\eta$ represent free parameters that were set to $0.9$ and $1.0$, respectively, in our baseline planet occurrence function (Equation~\ref{eqn:F21_occ_rate}). To test a variety of plausible stellar mass dependencies ranging from flat ($\eta\,{=}\,0$) to quadratic ($\eta\,{=}\,2$), we sampled $\eta$ from the uniform distribution $\mathcal{U}(0.0, 2.0)$. Similarly, to reflect our uncertainty in the normalization, we sampled $K$ from $\mathcal{U}(0.8, 1.0)$. Because we had already calculated the grid of searchable stars (left panel of Figure~\ref{fig:detection_grid}), introducing a new planet occurrence function simply involved multiplying by the new planet occurrence matrix and summing up the number of detections. This allowed us to perform simulations for $10{,}000$ random realizations of the planet occurrence function.

In the case of DR5 ($P_\mathrm{orb}\,{<}\,9.5$\,years and $N_\sigma\,{=}\,1.0$), the resulting distribution of the number of detected planets was approximately Gaussian, with a median of $130{,}000$ and a standard deviation of $25{,}000$ (relative uncertainty of about $20$\%). For DR4 ($P_\mathrm{orb}\,{<}\,4.0$\,years and $N_\sigma\,{=}\,1.5$), the median was $15{,}000$ and the standard deviation was $4{,}200$ (relative uncertainty of $30$\%). The uncertainties in the parameters of the \citet{Fulton2021} occurrence function and the uncertainties in the stellar mass dependence make comparable contributions to the total uncertainty.

The uncertainty is largest for detections around M-dwarfs, due to the variety of stellar mass extrapolations we considered. For DR5, our model predicts $31{,}000\,{\pm}\,18{,}000$ planet detections around stars with $M_\star\,{<}\,0.6\,M_\odot$, corresponding to a relative uncertainty of about $60$\%. For DR4, the analogous prediction is $5{,}600\,{\pm}\,3{,}300$ detected planets. The distributions are skewed in this case, with long tails in which there are many more M-dwarf detections than the median. The $95$\% confidence lower limit on the number of detections around M-dwarfs is $2{,}500$ for DR4 and $14{,}000$ for DR5.

In reality, given the approximations that were made when constructing the semi-analytic calculation, the results might not be accurate to within $30$\%. For example, there are systematic errors related to the simplified criteria for planet detection, the exponential-disk model for the spatial distribution of stars, and our neglect of stellar evolution. Below, we attempt to tackle these issues with more realistic models of the detection process and the stellar catalog.

\section{Simulated astrometric data}
\label{sec:astrom_data}

The semi-analytic calculation presented above relied on an idealized model for the detection process consisting of two criteria. The first criterion was that the \snr, the astrometric SNR per data point, must exceed a threshold value (Equation~\ref{eqn:SNR_criterion}). This criterion does not take into account the effects of the planet's orbital orientation, nor the actual number of data points collected by {\it Gaia}, which in turn depends on the star's ecliptic latitude (see \citealt{Holl2023b}). The second criterion was that the orbital period must be shorter than a certain maximum value. In reality, the detection probability falls continuously as the orbital period approaches and exceeds the observing baseline, as it becomes increasingly difficult to distinguish the star's orbital motion from its proper motion. To address these limitations, we generated and analyzed simulated {\it Gaia} astrometry with the help of the \texttt{Gaiamock} code \citep{ElBadry2024}.\footnote{Available at \url{https://github.com/kareemelbadry/gaiamock}}

The inputs required to generate synthetic astrometry are the star's right ascension (RA), declination (Dec), proper motion vector, mass, and $G$-band magnitude, along with the companion's mass, period, eccentricity, time of periapse, argument of periapse, inclination, and longitude of ascending node. \texttt{Gaiamock} generates synthetic along-scan astrometric data with a scanning law based on the {\it Gaia} Observation Forecast Tool\footnote{\url{https://gaia.esac.esa.int/gost/}} and the DR3 astrometric uncertainties reported by \citet{Holl2023a}. We used the \texttt{Gaiamock.predict\_astrometry\_luminous\_binary} function to simulate time-series astrometry, with the companion flux set equal to zero because planets are dark companions. To account for occasional bad data points, we randomly rejected $10$\% of the simulated field-of-view crossings. To account for systematic errors affecting bright stars ($G\,{<}\,13$; see \citealt{Lindegren2021, ElBadry2021}), we drew a random value in $\mu$as from the uniform distribution $\mathcal{U}(0,\,40)$ and added it to the 
default astrometric uncertainty $\sigma_\mathrm{fov}$. These and other features of the \texttt{Gaiamock} pipeline were motivated by comparisons between the code's predictions and the {\it Gaia} DR3 astrometric binary catalog \citep{Halbwachs2023} and the astrometric time series released for {\it Gaia} BH3 \citep{GaiaCollaboration2024}. We refer readers to \citet{ElBadry2024} for additional details.

\begin{figure*}
\centering
\includegraphics[width=0.95\textwidth]{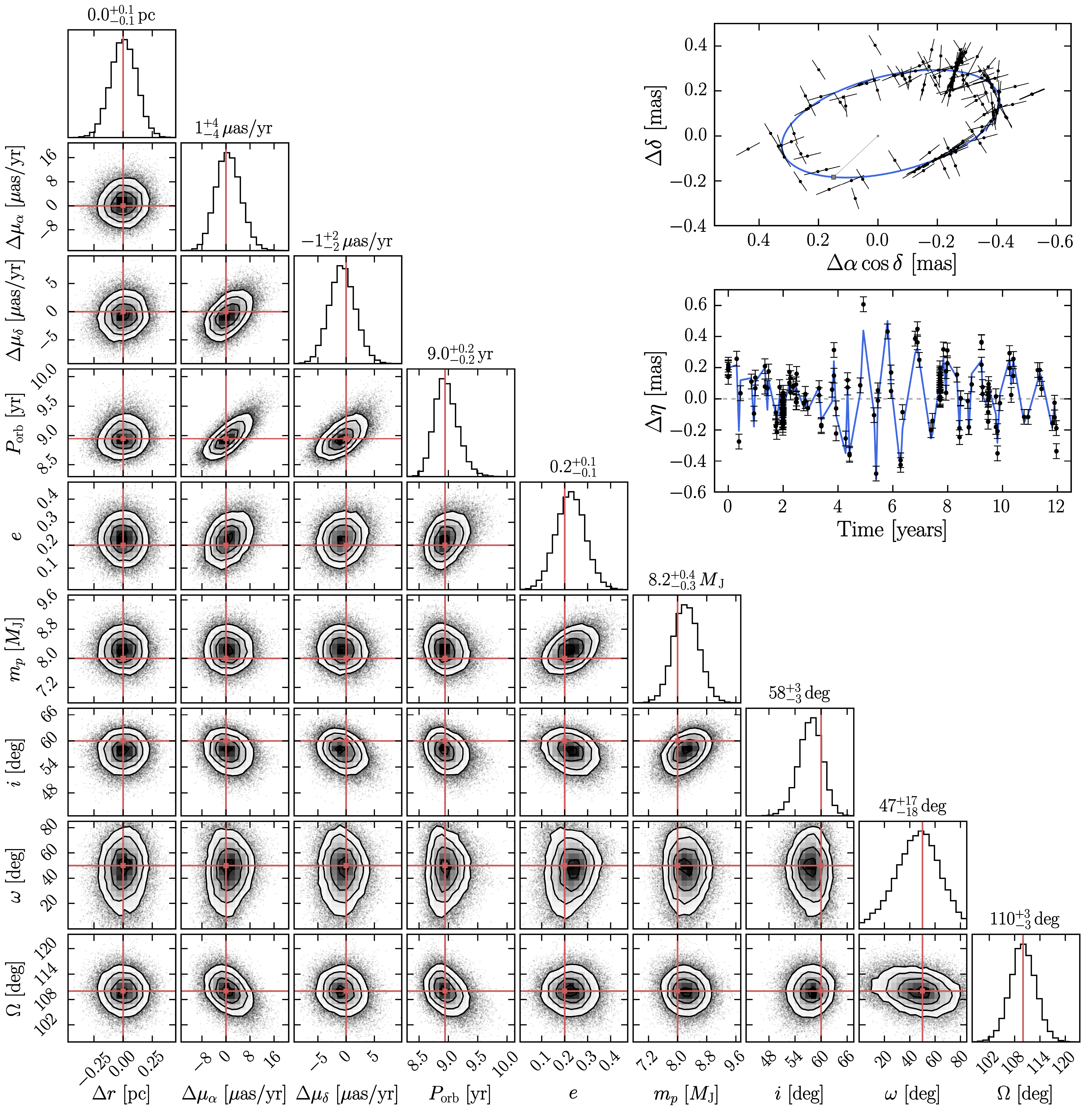}
\caption{Simulated {\it Gaia} DR5 astrometry for an $8~M_\mathrm{J}$ planet on a $4$~AU orbit around a $0.8~M_\odot$ star located $100$~pc away ($G\,{=}\,11$). The two panels in the upper right corner show the RA ($\alpha$) and Dec ($\delta$) measurements and the individual along-scan astrometric measurements ($\eta$). For both plots, we removed the best-fit proper motion and overplotted the best-fit planet model (blue). The rest of the figure is a corner plot showing the 2D posterior probability distributions for all model parameters except RA and Dec (which were omitted to keep the figure size manageable). Because the distance $r$ and proper motion components $\mu_\alpha$ and $\mu_\delta$ are tightly constrained, we have plotted their offsets from the true values ($\Delta r$, $\Delta \mu_\alpha$, and $\Delta \mu_\delta$). The true values of all parameters (red) were recovered accurately within the uncertainties. This system has \snr\,${\approx}\,7$; we expect that ${\sim}\,480$ planets in DR4 and ${\sim}\,2{,}300$ in DR5 will have \snr\,${>}\,7$ (see Figure~\ref{fig:det_vs_Nsigma}).}
\label{fig:corner_example}
\end{figure*}

As an illustrative example, Figure~\ref{fig:corner_example} shows mock {\it Gaia} DR5 astrometry for a $0.8~M_\odot$ star at a distance of $100$~pc that hosts a $8~M_\mathrm{J}$ planet on a $4$~AU orbit. Such a star has $G\,{=}\,11$, based on its distance and our adopted mass-magnitude relation (Figure~\ref{fig:mass-mag}). We chose plausible but arbitrary values for the sky coordinates, proper motion, and other orbital parameters (highlighted in red on the corner plot in Figure~\ref{fig:corner_example}). The maximum astrometric displacement caused by the planet is $382~\mu$as, well above the expected per-point precision of about $54~\mu$as for this star (see Figure~\ref{fig:astrom_unc}). Thus, in this example, the per-point signal-to-noise ratio is \snr\,${\approx}\,382/54\,{\approx}\,7$.

To simulate the detection process, we fitted the mock data with a no-companion model and with a one-companion model. The mock data consisted of the observation times $t_i$, the along-scan displacements $\eta_i$, the corresponding uncertainties $\sigma_{\eta_i}$, the scan angles $\psi_i$, and the parallax factors $\Pi_{\eta_i}$. In the no-companion model, the best-fit astrometric parameters (the RA $\Delta \alpha$ and Dec $\Delta \delta$ at a chosen reference time, the proper motion components $\mu_\alpha$ and $\mu_\delta$, and the parallax $\varpi$) were determined via linear regression. The no-companion model has a $\chi^2$ value of $2{,}288$ and $N_\mathrm{data}\,{-}\,N_\mathrm{params}\,{=}\,163$ degrees of freedom, indicating an unacceptable fit.

The one-companion model has 7 additional free parameters: the Thiele-Innes elements $A$, $B$, $F$, and $G$; the orbital period $P_\mathrm{orb}$, the eccentricity $e$, and the orbital phase $\phi_0\,{=}\,2\pi T_p/P_\mathrm{orb}$. As in the {\it Gaia} DR3 pipeline \citep{Halbwachs2023}, we determined the best-fit values of $P_\mathrm{orb}$, $e$, and $\phi_0$ parameters using non-linear optimization, and then found the best-fit values of the other parameters using linear regression. The non-linear optimization was performed with the adaptive simulated annealing algorithm implemented in \texttt{Gaiamock} (see Appendix~A of \citealt{ElBadry2024}). The Campbell elements $a_0$, $i$, $\omega$, and $\Omega$ were calculated from the Thiele-Innes elements as described in Appendix~A of \cite{Halbwachs2023}. As usual, $i$ is the inclination, $\omega$ is the argument of periapse, and $\Omega$ is the longitude of the ascending node. The planet mass was calculated from the orbital parameters as described in Appendix~\ref{sec:cubic_mass_eqn}.

For the mock data shown in Figure~\ref{fig:corner_example}, the one-companion model has a $\chi^2$ value of $280$ and $N_\mathrm{data}\,{-}\,N_\mathrm{params}\,{=}\,156$ degrees of freedom, indicating a more satisfactory fit, although with somewhat underestimated uncertainties (as expected for bright stars). The $\chi^2$ difference of $2{,}008$ between the no-companion and one-companion models indicates an overwhelming preference for the one-companion model. To determine the parameter uncertainties, we used the Markov Chain Monte Carlo (MCMC) algorithm implemented in the \texttt{emcee} code \citep{Foreman-Mackey2013}. For the MCMC analysis, we inflated the astrometric uncertainties by a factor of $\sqrt{280/156}\,{\approx}\,1.33$ so that the best-fit model had a reduced $\chi^2$ of unity. We used $100$ independent walkers, each taking $10{,}000$ steps, which was ${>}\,25$ times the autocorrelation length for each of the $12$ parameters. We discarded the first $5{,}000$ steps as burn-in. The resulting posterior is shown in the form of a corner plot in Figure~\ref{fig:corner_example}. The true parameters of the star and companion (highlighted in red) were recovered accurately. Based on the results shown in Figure~\ref{fig:det_vs_Nsigma}, the number of planets detected with \snr\,${>}\,7$ (i.e., at least as securely as in this example) is expected to be ${\sim}\,480$ in DR4 and ${\sim}\,2{,}300$ in DR5.

\subsection{Maximum orbital periods for DR4 and DR5}
\label{sec:max_Porb}

To investigate the detectability of planets with long orbital periods, we constructed synthetic planet-star systems using the procedure outlined below. First, we sampled stellar distances and sky positions assuming that the number density falls exponentially with vertical distance from the Galactic midplane, with a scale height of $300$~pc. Stars beyond $800$~pc were rejected and redrawn because planet detection is unlikely around such stars (see Section~\ref{sec:mock_catalog}). Then, we drew stellar masses from our model VMF$_\star$ (Figure~\ref{fig:VLF_VMF}) and assigned absolute magnitudes based on our mass-luminosity relation (Figure~\ref{fig:mass-mag}). Proper motions were chosen by sampling $v_\alpha$ and $v_\delta$ from independent random Gaussian distributions with a mean of zero and a standard deviation of $30$~km/s. We used our planet occurrence rate function (Equations~\ref{eqn:F21_occ_rate} and \ref{eqn:stellar_mass_dep}) to determine the occurrence rate of relevant planets (i.e., those with $a\,{\approx}\,0.1$\,--\,$7$\,AU and $m_{p}\,{\approx}\,0.3$\,--$13\,M_\mathrm{J}$). We randomly selected the appropriate number of planet-hosting stars, and sampled planetary $a$ and $m_p$ values based on Equations~\ref{eqn:F21_occ_rate} and \ref{eqn:stellar_mass_dep}. Eccentricities were sampled from a beta distribution with parameters $a\,{=}\,0.867$ and $b\,{=}\,3.03$ based on the results of \citet{Kipping2013}. Most of the resulting synthetic planets produce astrometric signals that are well below {\it Gaia}'s sensitivity level. For this reason, we excluded systems with \snr$\,{<}\,1.5$ and \snr$\,{<}\,1.0$ for our DR4 and DR5 experiments, respectively.

\begin{figure}
\centering
\includegraphics[width=0.475\textwidth]{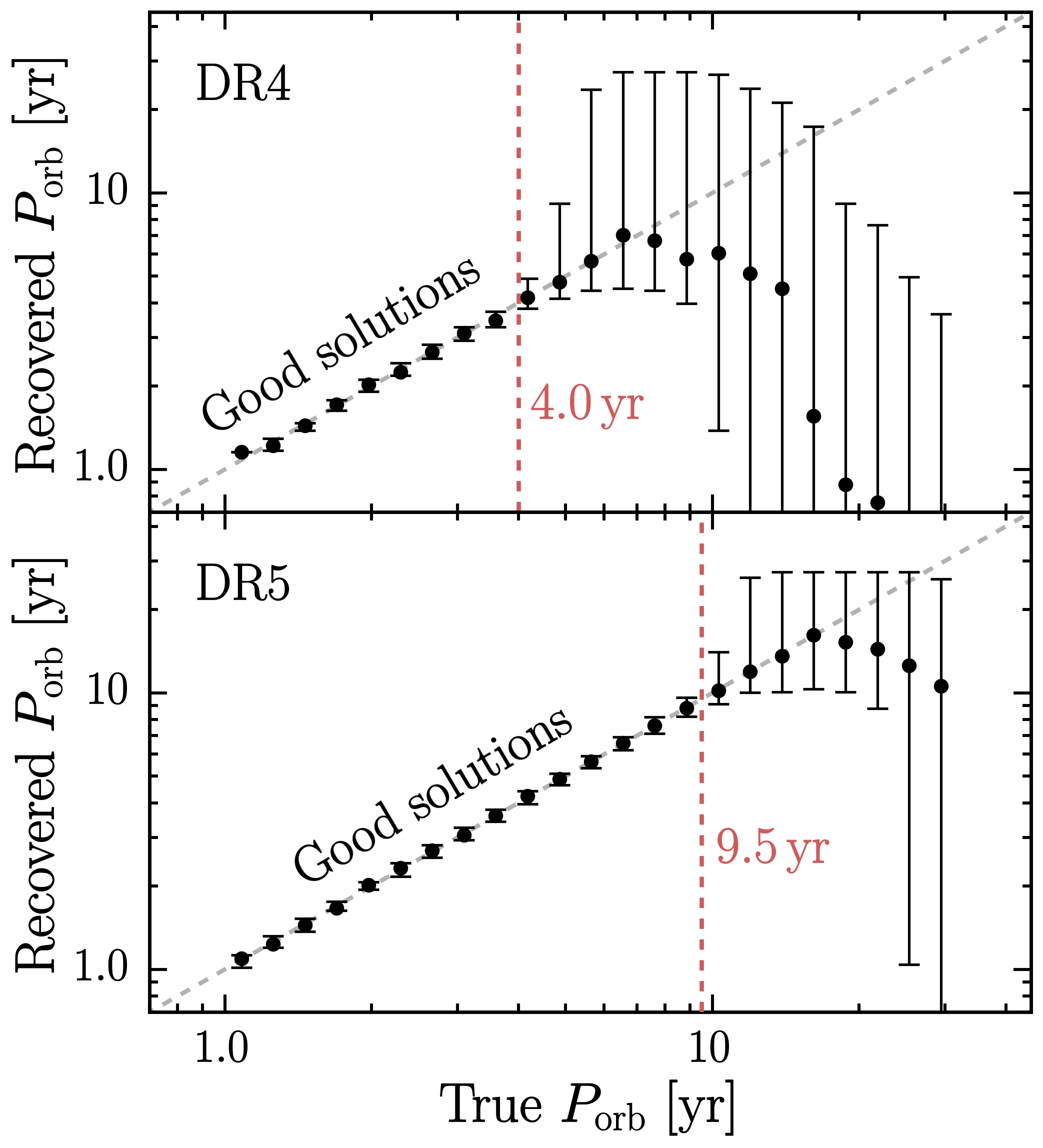}
\caption{Accuracy of fitted orbital periods, based on analyses of simulated {\it Gaia} astrometric time series. The top panel shows the recovered orbital period (i.e., the median and $1\sigma$ range) versus the true orbital period for simulated DR4 data. The bottom panel shows the same for simulated DR5 data. For the DR4 data, periods shorter than about $4.0$~years can be accurately recovered. For DR5, the limit is extended to about $9.5$~years.}
\label{fig:astrom_fitting_Porbs}
\end{figure}

Figure~\ref{fig:astrom_fitting_Porbs} shows the best-fit orbital period versus the true orbital period used to generate the simulated data, for DR4 (top) and DR5 (bottom). In creating this plot, we binned the data points according to their true periods, and in each bin, we plotted the median recovered period along with an error bar showing the standard deviation. In the DR5 simulations, orbital periods shorter than $5$~years were typically recovered with percent-level precision. As the period approaches the $10.5$~year baseline, the accuracy of the recovered orbital period worsens. At the nominal maximum period of $9.5$~years used in the semi-analytic model (Section~\ref{sec:analytic_model}), the mean relative error in $P_\mathrm{orb}$ is about $20$\%. As the period grows beyond $9.5$~years, the recovered periods quickly become unreliable. In the DR4 simulations, the results were similar except that an error of $20$\% was reached at $P_\mathrm{orb}\,{\approx}\,4.0$\,years instead of $9.5$~years. These $P_\mathrm{orb}$ thresholds correspond to $72$\% and $90$\% of the DR4 and DR5 baselines, respectively.

\subsection{Relationship between \snr, $\Delta \chi^2$, and measurement accuracy}
\label{sec:SNR_chi2_delta}

\begin{figure}
\centering
\includegraphics[width=0.475\textwidth]{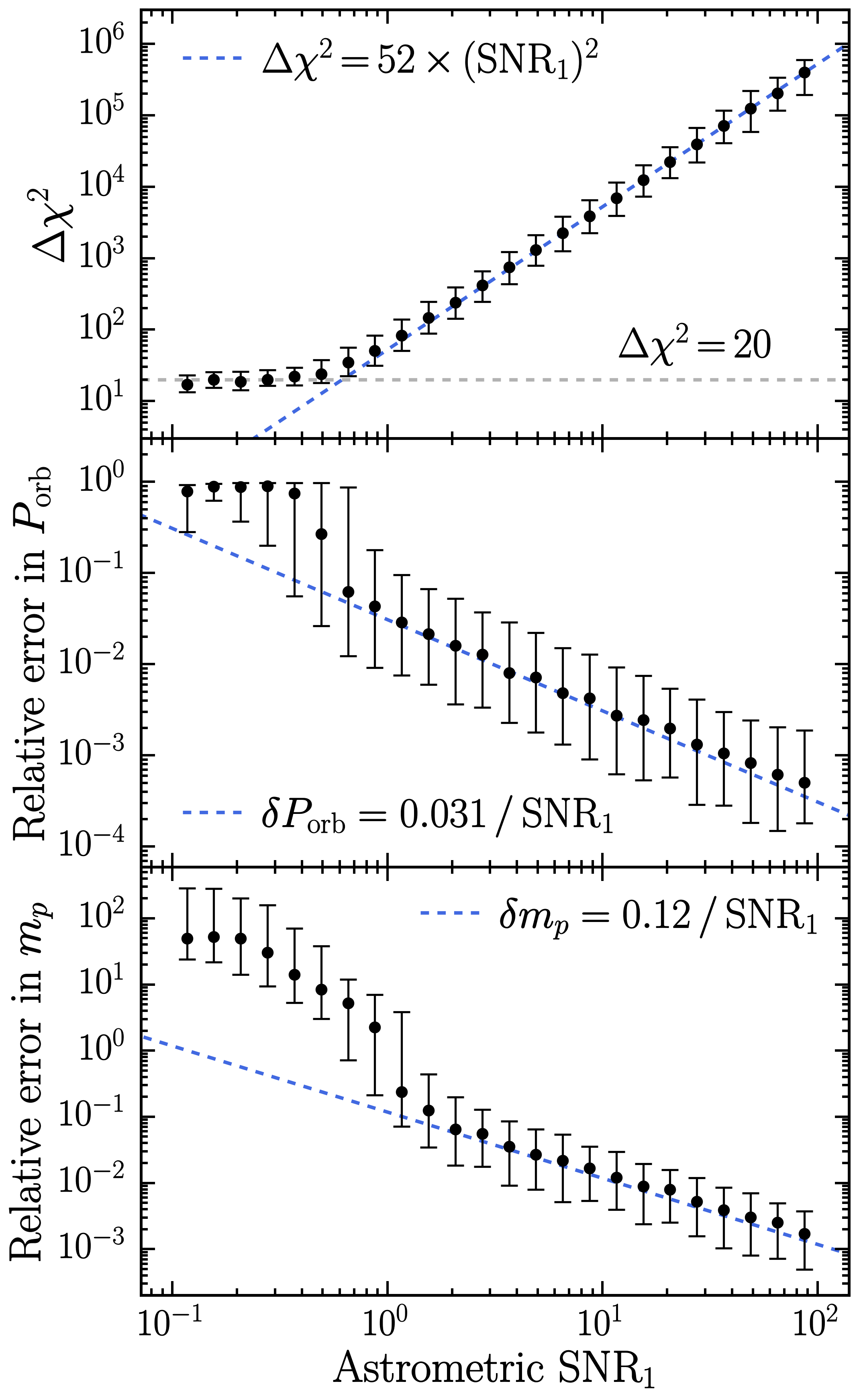}
\caption{Results of orbit-fitting experiments with simulated {\it Gaia} DR5 data. The points represent median values, and the error bars indicate $1\sigma$ ranges. Blue dashed lines show fits to the \snr$\,{>}\,2$ results. The top panel shows the $\Delta \chi^2$ between the best-fit one-companion and no-companion models, versus \snr, the signal-to-noise ratio of each astrometric data point (Equation~\ref{eqn:SNR_criterion}). When \snr\ is below about $0.5$, $\Delta \chi^2\,{\approx}\,20$. For \snr\,${\gtrsim}\,1$ there is a relatively tight relationship (within a factor of ${\sim}\,2$) between \snr\ and $\Delta \chi^2$. The middle panel shows the relative error in the recovered orbital period, $\delta P_\mathrm{orb}\,{=}\,1\,{-}\,P_\mathrm{recov}/P_\mathrm{true}$, versus \snr. $\delta P_\mathrm{orb}$ drops steadily with \snr. The relative error in the recovered planet mass versus \snr\ is shown in the bottom panel. Planet masses are accurately recovered when \snr$\,{\gtrsim}\,4$, and become increasingly accurate at larger \snr.}
\label{fig:astrom_fitting_SNRs}
\end{figure}

In the next set of orbit-fitting experiments, we explored the relationship between \snr\ and the strength of the evidence for a planetary companion, as well as the ability to measure the planet's parameters. Recall that \snr\ is the astrometric signal-to-noise ratio per data point, a simple and intuitive metric, but not a rigorous measure of the evidence for a planet detection based on all of the data. We quantified the evidence for a planet by computing the difference in $\chi^2$ between the best-fit no-companion and one-companion models. The top panel of Figure~\ref{fig:astrom_fitting_SNRs} shows the relationship between $\Delta \chi^2$ and \snr. Specifically, we plotted the median and standard deviation of the distribution of $\Delta\chi^2$ values within $24$ \snr\ bins. For this calculation, we imposed a maximum orbital period of $9.5$~years. We also modified the sampling of star-planet systems to ensure a nearly log-uniform distribution of \snr\ values by introducing an artificial ``acceptance probability'' that increases with \snr. Otherwise, if we had initialized the star-planet systems as described in Section~\ref{sec:max_Porb}, the low-\snr\ systems would have vastly outnumbered the high-\snr\ systems, making it computationally expensive to generate enough high-\snr\ systems for the summary statistics to converge.

Systems with \snr\,${\lesssim}\,0.5$ typically favor the one-companion model over the no-companion model by $\Delta \chi^2\,{\approx}\,20$ (Figure~\ref{fig:astrom_fitting_SNRs}), which is not high enough to justify the inclusion of the $7$ extra free parameters. The Bayesian information criterion (BIC) statistic, $\mathrm{BIC}\,{=}\,\chi^2\,{+}\,k\ln(n)$, is often used to penalize models with more free parameters. In this equation, $k$ is the number of free parameters and $n$ is the number of data points \citep{Schwarz1978}. In our DR5 simulations, $\Delta \chi^2\,{=}\,20$ corresponds to $\Delta \mathrm{BIC}\,{\approx}\,-16$. The negative value indicates a preference for the no-companion model.

As \snr\ is increased, $\Delta \chi^2$ grows approximately as (\snr)$^2$, as expected for independent homoskedastic measurement uncertainties. The blue dashed line in the top panel of Figure~\ref{fig:astrom_fitting_SNRs} is the quadratic function that best fits the system-by-system results with \snr$\,{>}\,2$. The quadratic function provides a good description of the data, with a dispersion of approximately a factor of two. The nominal \snr\ threshold from Section~\ref{sec:analytic_model}, $N_\sigma\,{=}\,1.0$, corresponds to $\Delta \chi^2\,{\approx}\,50$ and $\Delta \mathrm{BIC}\,{\approx}\,14$, representing modest evidence for a companion.

To study how accurately the parameters of detected planets will be constrained, we recorded the best-fit parameters after each orbit-fitting experiment. The middle panel of Figure~\ref{fig:astrom_fitting_SNRs} shows the relative error in the orbital period ($\delta P_\mathrm{orb}\,{=}\,1\,{-}\,P_\mathrm{recov}/P_\mathrm{true}$) versus \snr. For systems with \snr\,${\lesssim}\,0.5$, orbital periods are recovered poorly, with $\delta P_\mathrm{orb}\,{\approx}\,1$. Once \snr$\,{\approx}\,1.0$, about $90$\% of systems have orbital periods recovered with better than $20$\% accuracy. The median accuracy increases steadily with \snr, reaching ${\sim}\,10^{-3}$ for systems with \snr$\,{\approx}\,100$. We fitted a function that scales inversely with \snr\ for systems with \snr$\,{>}\,2$. The resulting relation accurately reproduces the overall trend of median $\delta P_\mathrm{orb}$ versus \snr, with a dispersion of about a factor of four.

Turning now to planet mass measurements, the bottom panel of Figure~\ref{fig:astrom_fitting_SNRs} shows the relative error in the recovered planet masses versus \snr. In order to reduce the error in the recovered mass below $20$\% at least $90$\% of the time, \snr\ must be at least $2.5$, several times higher than is needed to attain $20$\% accuracy in the orbital period. The accuracy of the recovered masses also scales inversely with \snr, with a dispersion of a factor of three about the best-fit function.

Although the results are not plotted, we performed a similar analysis of the accuracy of the recovered orbital inclinations and eccentricities. The results for recovered inclinations are similar to the analogous results for mass. Achieving an error below $20$\% in inclination for more than $90$\% of systems requires \snr$\,{\approx}\,2.5$, and the scatter around the best-fit inverse function is about a factor of three. In contrast, measuring accurate eccentricities requires substantially larger \snr\ values. The median error in the recovered eccentricities only dropped below $20$\% for systems with \snr$\,{\approx}\,9$. The fraction of systems with eccentricities measured to within $20$\% only reached $90$\% when \snr$\,{\approx}\,50$. The overall trends are summarized below:
\begin{align}
\label{eqn:DR5_chi2_SNR}
\Delta \chi^2 &\approx 52\,{\times}\,(\mathrm{SNR}_1)^2 \\
\label{eqn:DR5_Porb_SNR}
\delta P_\mathrm{orb} &\approx 0.031\,{/}\,\mathrm{SNR_1} \\
\label{eqn:DR5_mp_SNR}
\delta m_p &\approx 0.12\,{/}\,\mathrm{SNR_1} \\
\label{eqn:DR5_inc_SNR}
\delta i &\approx 0.11\,{/}\,\mathrm{SNR_1} \\
\label{eqn:DR5_ecc_SNR}
\delta e &\approx1.4\,{/}\,\mathrm{SNR_1}~\mathrm{({\it Gaia}~DR5)}
\end{align}

We repeated the orbit-fitting experiments described above with simulated DR4 data, restricting our attention in this case to planets with orbital periods shorter than $4.0$~years. We found the $\Delta \chi^2$ and relative errors in orbital elements and planet mass to have similar scalings with \snr\ but different coefficients:
\begin{align}
\label{eqn:DR4_chi2_SNR}
\Delta \chi^2 &\approx 22\,{\times}\,(\mathrm{SNR}_1)^2 \\
\label{eqn:DR4_Porb_SNR}
\delta P_\mathrm{orb} &\approx 0.046\,{/}\,\mathrm{SNR_1} \\
\label{eqn:DR4_mp_SNR}
\delta m_p &\approx 0.20\,{/}\,\mathrm{SNR_1} \\
\label{eqn:DR4_inc_SNR}
\delta i &\approx 0.18\,{/}\,\mathrm{SNR_1} \\
\label{eqn:DR4_ecc_SNR}
\delta e &\approx 2.3\,{/}\,\mathrm{SNR_1}~\mathrm{({\it Gaia}~DR4)}
\end{align}
Because the recovered masses, inclinations, and eccentricities were often unreliable for systems with \snr$\,{\lesssim}\,2$, we restricted the fits to systems with \snr$\,{>}\,3$. For DR4, requiring $\Delta \chi^2\,{>}\,50$ corresponds to \snr$\,{\gtrsim}\,1.5$. At a fixed \snr, DR5 provides constraints on orbital parameters that are about $1.7$ times smaller than those provided by DR4. Measuring $P_\mathrm{orb}$, $m_p$, and $i$ with $20$\% accuracy for $90$\% of systems requires \snr\ values larger than about $1.5$, $4$, and $4$, respectively.

The semi-analytic model from Section~\ref{sec:analytic_model} predicts that ${\sim}\,21{,}000$ planets will be detected in DR5 with \snr$\,{>}\,2.5$ and ${\sim}\,1{,}800$ planets will be detected in DR4 with \snr$\,{>}\,4$. We expect the masses of these planets to be measured relatively securely. Because some lower \snr\ planets will also have their masses constrained, these predictions are probably underestimates, and are improved below.

\section{Mock exoplanet catalogs}
\label{sec:mock_catalog}

Armed with the results from the semi-analytic model in Section~\ref{sec:analytic_model} and the orbit-fitting experiments described in Section~\ref{sec:astrom_data}, we are ready to construct mock {\it Gaia} exoplanet catalogs for DR4 and DR5. We envision multiple uses for these catalogs. First, we expect that the catalogs will help set expectations for the planet yield in the coming data releases, including the quality of the constraints placed on parameters of interest. Second, having the properties of the relevant stars in hand before the {\it Gaia} data are released may help with planning observational follow-up campaigns. Third, once the real {\it Gaia} data are available, we hope that these catalogs will serve as a useful point of comparison, with the potential to improve our understanding of {\it Gaia}'s systematics and giant-planet occurrence.

\subsection{Observationally relevant stars}
\label{sec:obs_rel_stars}

Our starting point is the {\it Gaia} DR3 catalog, which contains data for nearby stars that is complete to at least $G\,{\approx}\,18$ \citep{CantatGaudin2023}. The first challenge is the large scale of the dataset. A staggering ${\sim}\,1.4$~billion stars have {\it Gaia}-measured positions, parallaxes, proper motions, and photometry. Planets are detectable around a relatively small subset of these stars. To restrict attention to these observationally relevant stars, we used our semi-analytic model to calculate the maximum distance out to which an ``ideal planet'' -- a planet of mass $13~M_{\rm J}$ on a $9.5$-year orbit -- could be detected, as a function of absolute magnitude (see Equations~\ref{eqn:volume-integral},~\ref{eqn:volume-integral-2} and Figure~\ref{fig:rmax_Gmax}). This calculation employs the relation between mass and absolute magnitude shown in Figure~\ref{fig:mass-mag}. We adopted a weak detection threshold, $N_\sigma\,{=}\,0.5$, to avoid discarding any stars for which planet detections are possible. For bright stars ($M_G\,{\lesssim}\,2.8$), the maximum distance decreases with luminosity, while for fainter stars, the maximum distance increases with luminosity. The maximum search distances for bright and faint star scalings are well described by the fitting functions 
\begin{align}
\label{eqn:obs_rel_stars}
r_\mathrm{max} &\approx 1600\,\mathrm{pc} \times10^{-0.069 (M_G - 2.8)}~(\mathrm{for}~M_G \lesssim 2.8) \\
r_\mathrm{max} &\approx 1600\,\mathrm{pc} \times 10^{0.093 (M_G - 2.8)}~ (\mathrm{for}~M_G \gtrsim 2.8)~.
\end{align}
We queried the {\it Gaia} DR3 source catalog for stars having $r\,{<}\,r_{\rm max}(M_G)$, where $M_G$ was calculated under the assumption that extinction is negligible.

Then, we used {\it Gaia} photometric colors to restrict our attention to main-sequence stars. Specifically, we retained all stars satisfying
\begin{equation}
\label{eqn:main_seq_cut}
     0.4 < M_G - 2.9(G_{BP} - G_{RP}) < 4.5~,
\end{equation}
where $G_{BP}$ and $G_{RP}$ are the apparent magnitudes in the {\it Gaia} ``blue photometer'' and ``red photometer'' bands (see \citealt{DeAngeli2023}). The slope of $2.9$ and the threshold values of $0.4$ and $4.5$ were chosen to isolate the main-sequence strip of the color-magnitude diagram. Our results for the total number of planet detections do not depend sensitively on the details of this cut; modifications mainly affect the number of high-mass stars in the sample, for which planet detections are rare. In total, $38.3$~million stars satisfied our $M_G$-dependent distance cut and our main-sequence criteria.

To assign planets and simulate astrometric data, we needed estimates for the stars' masses. The {\it Gaia} team provided mass estimates for $140$~million stars based on low-resolution spectra obtained with the Radial Velocity Spectrometer (RVS) instrument \citep{Creevey2023}. However, of the $38.3$~million observationally relevant main-sequence stars, only $19.5$~million ($51$\%) were assigned masses in the DR3 astrophysical parameters table. Most of the stars with missing masses are low-mass stars ($M_\star\,{\lesssim}\,0.5\,M_\odot$; \citealt{Creevey2023}). We decided to use the masses provided by the {\it Gaia} team when they were available, and to otherwise assign masses based on the mass-magnitude relation shown in Figure~\ref{fig:mass-mag}. As a consistency check, we compared the masses predicted by the mass-luminosity relation to the {\it Gaia}-reported masses for the $19.5$~million relevant stars with mass measurements. The median fractional difference was $6.7$\%. Such a comparison was not possible for low-mass stars (${<}\,0.5\,M_\odot$) because they lack {\it Gaia}-reported masses, but we expect the mass-luminosity relation to be fairly accurate in this regime because the effects of stellar evolution are minimal.

\subsection{Injection-recovery setup}
\label{sec:inject_recover_setup}

With the set of relevant {\it Gaia} stars in hand, we assigned planets and simulated {\it Gaia} observations. For each star, we used the occurrence function described in Section~\ref{sec:planet_occ} to calculate the probability of hosting a giant planet with a mass between $0.3$ and $13$~$M_\mathrm{J}$ and an orbital semi-major axis between $0.1$ and $7$~AU. We assigned planets accordingly, sampling planet masses and semi-major axes using Equation~\ref{eqn:F21_occ_rate}. Eccentricities were drawn from a beta distribution with parameters $a\,{=}\,0.867$ and $b\,{=}\,3.03$ \citep{Kipping2013}. We assigned at most one planet per star, a simplification discussed further in Section~\ref{sec:complications}. Then, we generated mock astrometry for {\it Gaia} DR4 and DR5 using the \texttt{Gaiamock} package, as described in Section~\ref{sec:astrom_data}. For each star, we used the best-fit RA, Dec, proper motion, and $G$-band magnitude reported in the {\it Gaia} source catalog.

Many planets assigned via this process lie well below {\it Gaia}'s detection sensitivity. To reduce the computational expense of orbit fitting, we discarded systems with an \snr\ below $0.5$ or an orbital period longer than $7$~years for DR4 or $14$~years for DR5. These choices were motivated by the results presented in Section~\ref{sec:astrom_data}, which indicated that there is little hope of detecting such planets. For all the systems surviving these cuts, we fitted the simulated along-scan time-series astrometric data with a no-companion model and a one-companion model (as described in Section~\ref{sec:astrom_data}). Whenever the $\Delta \chi^2$ between the best-fit one- and no-companion models exceeded $50$, we carried out an MCMC analysis to determine the uncertainties in the companion's parameters. We used $100$ independent walkers, each of which took $10{,}000$ steps, the first $5{,}000$ of which were discarded (as in Figure~\ref{fig:corner_example}). Before the MCMC analysis, we inflated the astrometric uncertainties so that the best-fit model had a reduced $\chi^2$ of unity, to avoid underestimating the parameter uncertainties. We chose to limit attention to systems with $\Delta \chi^2\,{>}\,50$, which corresponds to $\Delta \mathrm{BIC}\,{\gtrsim}\,19$ and $14$ for DR4 and DR5 data, respectively, indicating at least modest evidence for a planet. We deemed a planet to be detected when three criteria were met:
\begin{enumerate}
\item The best-fit one-companion model is favored over the no-companion model by $\Delta \chi^2\,{>}\,50$.
\item The best-fit planet mass is lower than $13\,M_\mathrm{J}$.
\item The orbital period is at least moderately constrained. Specifically, we required $P_\mathrm{84th}/P_\mathrm{16th}\,{<}\,1.5$, where $P_\mathrm{xth}$ is the $x$th percentile of the marginalized posterior probability distribution for the orbital period.
\end{enumerate}
Of course, these criteria are somewhat arbitrary, including the $13~M_\mathrm{J}$ limit. In practice, the chosen criteria will depend on the goals of the investigation. For this reason, we also report below how the number of detected planets varies with the detection thresholds.

\begin{figure*}
\centering
\includegraphics[width=0.95\textwidth]{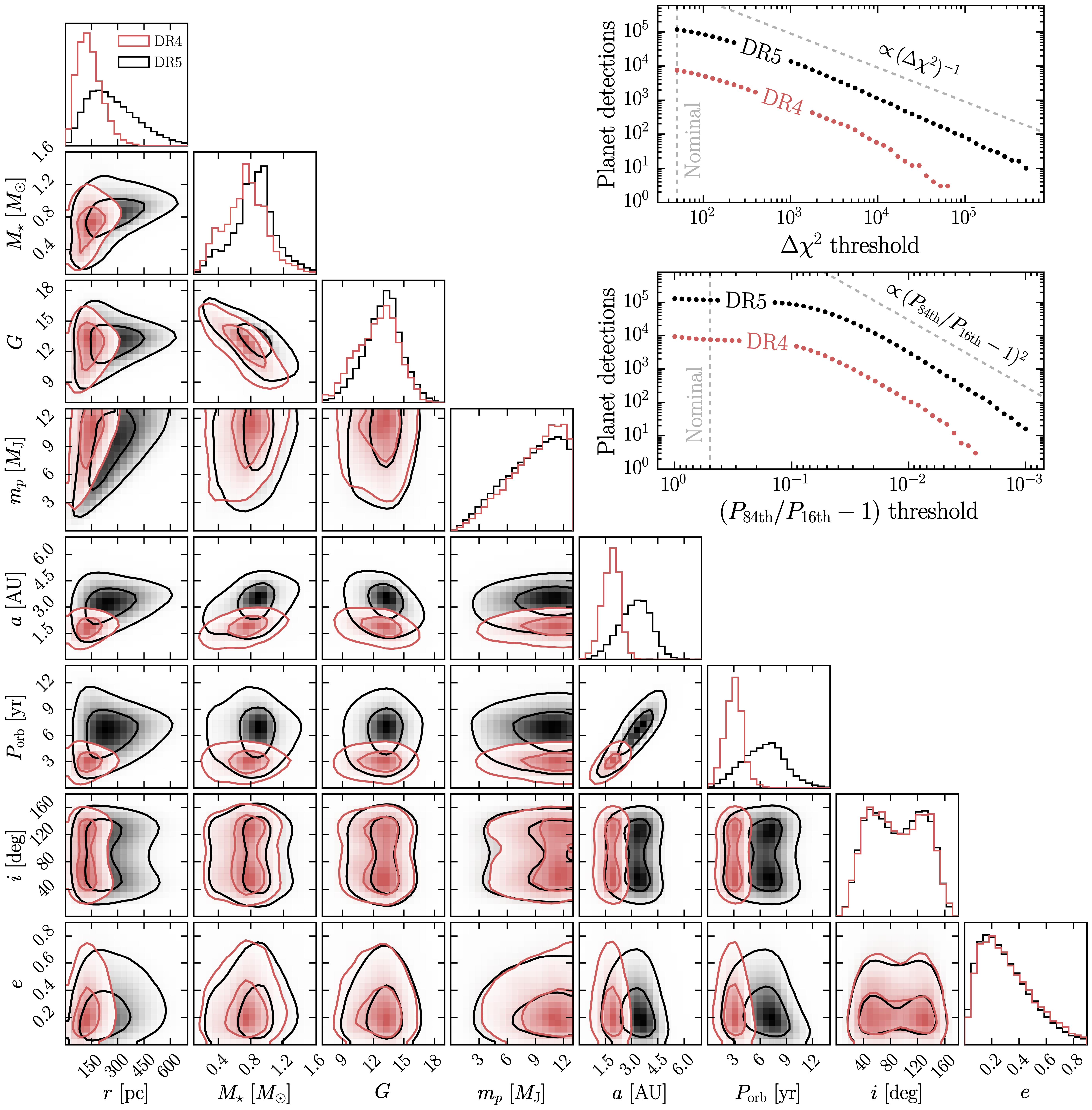}
\caption{Recovered properties for the $7{,}500$ mock {\it Gaia} DR4 detections (red) and the $120{,}000$ DR5 detections (black). Contours show the $1\sigma$ and $2\sigma$ ranges in each 2D distribution, and the histograms show the 1D distributions. The plots in the top right corner show how the number of detections depends on the $\Delta \chi^2$ threshold and the orbital period precision threshold (see Section~\ref{sec:inject_recover_results}). Most detections will be super-Jupiters on few-AU orbits around GKM-type stars within a few hundred parsecs. The longer timespan of the DR5 data will dramatically increase {\it Gaia}'s search volume, resulting in ${\sim}\,16$ times as many detections as in DR4. Only a subset of the detected planets will have well-constrained masses; we predict that $1{,}900\,{\pm}\,540$ and $38{,}000\,{\pm}\,7{,}300$ planets will have masses and periods measured to within $20$\% in DR4 and DR5, respectively (see Section~\ref{sec:inject_recover_results}).}
\label{fig:mock_catalog}
\end{figure*}

\subsection{Injection-recovery results}
\label{sec:inject_recover_results}

A total of $120{,}000$ planets in the DR5 simulation passed our nominal detection criteria, as compared to $7{,}500$ planets in the analogous DR4 simulation. Figure~\ref{fig:mock_catalog} is a corner plot showing the distribution of recovered properties for the detected planets and their host stars in the mock DR4 and DR5 catalogs.

In agreement with the semi-analytic calculation, nearly all of the detected planets are super-Jupiters (${\gtrsim}\,3\,M_\mathrm{J}$) on few-AU orbits ($1$\,--\,$3$\,AU in DR4 and $2$\,--\,$5$\,AU in DR5). Most DR4 planets orbit GKM-type stars ($M_\star\,{\approx}\,0.4$\,--\,$1.0\,M_\odot$) at distances of $50$\,--\,$250$\,pc ($G\,{\approx}\,9$\,--\,$15$). Fewer than $10$\% of the DR4 planets orbit stars beyond $250$~pc. About $30$\% of the DR4 planets orbit M-dwarfs ($M_\star\,{<}\,0.6\,M_\odot$).

The longer timespan of DR5 enables the discovery of planets at larger orbital distances ($3$\,--\,$5$\,AU), and since such planets produce larger-amplitude astrometric signals, the result is a vast expansion of the search volume and a corresponding enlargement in the number of detected planets. The DR5 planets have host stars that span a somewhat wider range of masses ($M_\star\,{\approx}\,0.5$\,--\,$1.2\,M_\odot$) and a wider range of distances of $100$\,--\,$500$\,pc ($G\,{\approx}\,10$\,--\,$16$). Fewer than $10$\% of the DR5 planets orbit stars beyond $500$~pc. About $17$\% of the DR5 planet detections involve M-dwarfs, a smaller fraction than for DR4 --- but in absolute terms, the M-dwarf planets in DR5 outnumber those in DR4 by a factor of $9$.

Few DR4 planets have periods exceeding $4.0$~years, and few DR5 planets have periods exceeding $9.5$~years, justifying the choices of maximum periods that were imposed in our semi-analytic model. The primary known source of uncertainty in the number of planet detections arises from the uncertainties in the planet occurrence function, which we found in Section~\ref{sec:unc_planet_occ} to be $20$\,--\,$30$\%. Reusing the relative uncertainties, we predict $7{,}500\,{\pm}\,2{,}100$ planet detections with DR4 data, and $120{,}000\,{\pm}\,22{,}000$ planet detections with DR5 data. These detection numbers are broadly consistent with those of the semi-analytic model. It is possible, of course, that there are yet-unknown systematics in {\it Gaia}'s DR4/DR5 astrometry, which could lower the planet yield. Conversely, improvements in systematic mitigation in the coming data releases may improve {\it Gaia}'s astrometric precision for bright stars ($\sigma_0$), which would increase the planet yield.

The distribution of parameters depicted in Figure~\ref{fig:mock_catalog} is strongly sculpted by observational biases. All of the detected low-mass planets ($m_p\,{<}\,3\,M_\mathrm{J}$) orbit stars with $r\,{\lesssim}\,200$\,pc. Similarly, detections around low-mass stars ($M_\star\,{<}\,0.4\,M_\odot$) only occur within ${\lesssim}\,250$\,pc. As distance increases, the fraction of detections involving massive planets and stars also increases. Within $150$~pc, the host stars of detected planets span a wide range of apparent magnitudes from $G\,{\approx}\,9$\,--\,$17$. For stars more than $400$~pc away, detections are more concentrated around stars with $G\,{\approx}\,13$. Observational bias also causes the mass distribution of detected planets to be top-heavy, even though the intrinsic mass distribution is bottom-heavy.

Some other biases are more subtle. Even though the simulated planets have random orbital orientations, there is a mild deficit of detected planets with nearly edge-on orbits ($i\,{\approx}\,90^\circ$), because astrometry provides less information about their orbits; one dimension is lost to the sky projection (see \citealt{GaiaCollaboration2023, Makarov2025}). Additionally, relatively few detected planets have $e\,{\approx}\,0$, even though many simulated planets have low-eccentricity orbits. The reason is that {\it Gaia}'s constraints on the eccentricity are generally poor. Since zero is the minimum permissible value of $e$, most planets with $e\,{\approx}\,0$ end up with best-fit eccentricities of $e\,{\approx}\,0.2$ due to noise. This is sometimes referred to as the \citet{Lucy&Sweeney1971} bias. Host stars are distributed roughly uniformly on the sky, with mild biases imprinted by {\it Gaia}'s scanning law.

The plot in the top right corner of Figure~\ref{fig:mock_catalog} shows the total number of planet detections versus the $\Delta \chi^2$ threshold. The two other detection criteria (best-fit $m_p\,{<}\,13\,M_{\rm J}$ and $P_\mathrm{84th}/P_\mathrm{16th}\,{<}\,1.5$) were unchanged in this experiment. The number of detections scales approximately as ${\propto}\,(\Delta \chi^2)^{-1}$, as expected from the previously noted trends $N_\mathrm{detected}\,{\propto}\,N_\sigma^2$ (Figure~\ref{fig:det_vs_Nsigma}) and $\Delta \chi^2\,{\propto}\,$\snr$^2$ (Figure~\ref{fig:astrom_fitting_SNRs}). The dependence becomes less steep for $\Delta \chi^2\,{\lesssim}\,200$, where orbital periods and masses are less well constrained, making the effects of the other two criteria more significant. Figure~\ref{fig:mock_catalog} also includes a plot of the planet yield versus the threshold for orbital period precision, quantified as $P_\mathrm{84th}/P_\mathrm{16th}$. The number of planet detections drops sharply with $P_\mathrm{84th}/P_\mathrm{16th}$ for $P_\mathrm{84th}/P_\mathrm{16th}\,{\lesssim}\,1.1$. The dependence is less steep for $P_\mathrm{84th}/P_\mathrm{16th}\,{\gtrsim}\,1.1$, but the differences are still meaningful. For instance, adopting the threshold $P_\mathrm{84th}/P_\mathrm{16th}\,{<}\,2$ increases the number of DR4 detections by $24$\% and the number of DR5 detections by $11$\%. Allowing for larger values of $P_\mathrm{84th}/P_\mathrm{16th}$ enables the detection of longer-period planets, eventually making our $P_\mathrm{orb}\,{<}\,7$\,years (DR4) and $P_\mathrm{orb}\,{<}\,14$\,years (DR5) conditions restrictive.

The predicted yield also depends on the requirement that the best-fit mass be lower than $13\,M_{\rm J}$. Our results can be extrapolated to consider larger mass cutoffs, although we note that the occurrence rate of such objects is relatively uncertain (see \citealt{Holl2022}). Most of the detected planets shown in Figure~\ref{fig:mock_catalog} have poor mass constraints, leaving open the possibility they are brown dwarfs or, in some cases, low-mass stars. As a stricter detection criterion, we also tallied planets with masses and orbital periods constrained to within $20$\% (i.e., $P_\mathrm{84th}/P_\mathrm{16th}\,{<}\,1.2$ and $m_\mathrm{84th}/m_\mathrm{16th}\,{<}\,1.2$). Our DR4 catalog includes $1{,}900\,{\pm}\,540$ such planets, and our DR5 catalog includes $38{,}000\,{\pm}\,7{,}300$. The properties of these systems resemble those in Figure~\ref{fig:mock_catalog}, although they tend to orbit more nearby stars. The number of planets with accurate mass measurements agrees fairly well with the semi-analytic model, which predicted that ${\sim}\,1{,}800$ planets in DR4 and ${\sim}\,21{,}000$ in DR5 would have a sufficiently large \snr\ to have their masses measured (Section~\ref{sec:SNR_chi2_delta}).

Our mock {\it Gaia} exoplanet catalogs for DR4 and DR5, including true parameters, best-fit parameters, and MCMC constraints, are publicly available on GitHub.\footnote{\url{https://github.com/CalebLammers/GaiaForecasts}}

\begin{figure*}
\centering
\includegraphics[width=0.95\textwidth]{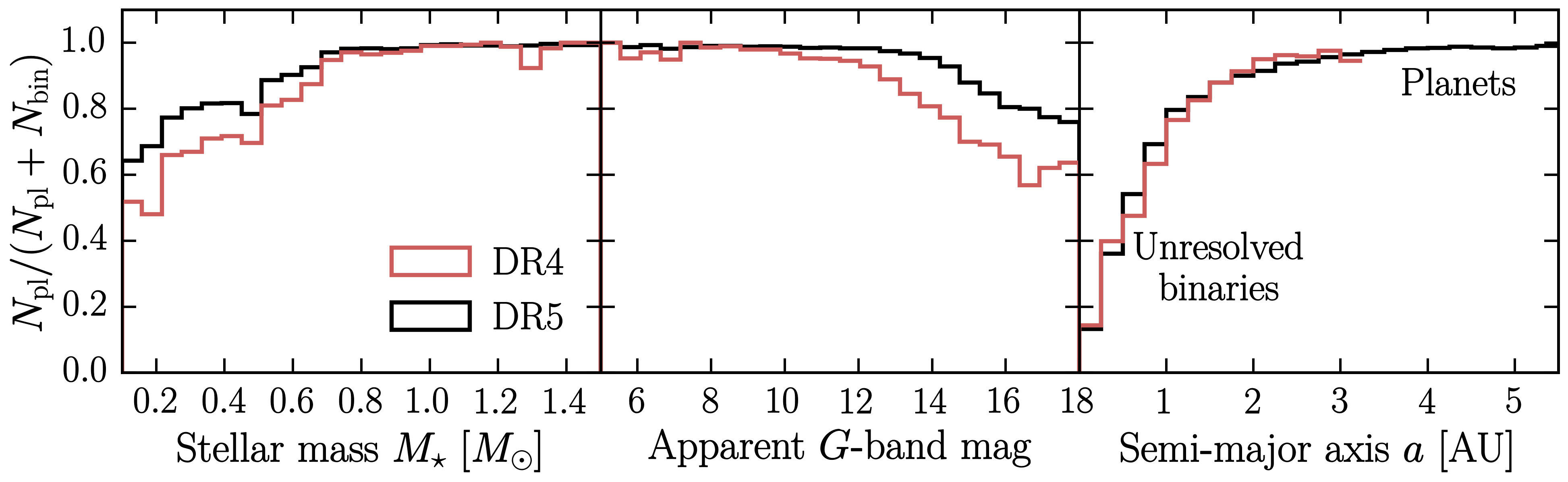}
\caption{The expected fraction of astrometric planet candidates
that are genuine planets (instead of unresolved binaries) as a function of apparent stellar mass $M_\star$, $G$-band magnitude, and semi-major axis $a$. For most values of $M_\star$, $G$, and $a$, our simulations predict that ${\gtrsim}\,80$\% of candidates will be genuine planets. The expected fraction of candidates that are unresolved binaries is largest for low-mass stars, faint stars, and small orbital separations, in which case it can exceed $50$\%.}
\label{fig:planet_binary_hists}
\end{figure*}

\section{Binary stars as planet impostors}
\label{sec:binary_stars}

Even when {\it Gaia} securely detects astrometric motion with a low enough amplitude to be compatible with a planetary-mass companion, a planet is not the only possible explanation. It could also be an unresolved binary star whose members have nearly equal $G$-band luminosities. The partial cancellation of the opposite motion of two stars can cause an unresolved binary to exhibit low-amplitude motion. If the cancellation is nearly complete, the resulting astrometric motion can mimic that of a single star with a dark planetary-mass companion. Indeed, unresolved near-twin binaries have been the primary false positive that has been encountered in ongoing investigations of the {\it Gaia} DR3 exoplanet candidates. More than half of the candidates that have been followed up with RV observations turned out to be unresolved nearly twin binaries \citep{Marcussen2023, Stefansson2025}.

How much will the DR4 and DR5 exoplanet catalogs be contaminated by false positives? It is unclear whether the incidence of unresolved binaries as planet impostors will be comparable to that of the DR3 catalog, given the strong observational biases as well as the complex process used to select the DR3 exoplanet candidates (see \citealt{GaiaCollaboration2023}). To size up the problem and help the community prepare for the follow-up effort, we tackled this question with additional simulations.

\subsection{Simulated binary population}
\label{sec:sim_binaries}

We generated a population of synthetic binary stars using a similar approach to that of \citet{ElBadry2024}. With the help of the \texttt{COSMIC} binary population synthesis code \citep{Breivik2020}, we sampled binary properties from the joint mass-ratio orbital-period model of \citet{Moe&DiStefano2017}, assuming a \citet{Kroupa2001} initial mass function for the primaries. We adopted a constant star formation rate over $12$~Gyr, and drew [Fe/H] values randomly from a Gaussian distribution $\mathcal{N}(0, 0.2)$. The two components of a given binary were assumed to have the same age and [Fe/H]. We stopped generating binaries once the number of stellar systems (singles plus binaries) exceeded $860$~million, the approximate number of systems within $2{,}000$~pc of the Sun \citep{ElBadry2024}. The limiting distance of $2{,}000$~pc was chosen because the vast majority of {\it Gaia}'s DR3 astrometric binaries are located within this distance \citep{Halbwachs2023}. The $G$, $G_{BP}$, and $G_{RP}$ magnitudes were assigned to each star based on its mass, age, and [Fe/H] using the MIST models \citep{Choi2016} that were accessed with the \texttt{isochrones} code \citep{Morton2015}. Binaries containing stars whose ages exceed the maximum age predicted by the MIST models were discarded. We neglected binary interactions, which are not expected to affect the main-sequence binaries in AU-scale orbits that make up the {\it Gaia} astrometric binary population.

By default, \texttt{COSMIC} does not assign positions to the simulated binaries. To determine which binaries would be unresolved by {\it Gaia}, we assigned 3D positions by sampling from an exponential thin-disk model with a vertical scale height of $300$~pc and a radial scale length of $2{,}500$~pc. We rejected and re-drew stars located beyond $2{,}000$~pc. Orbits were assigned random orientations, with $\cos i$ sampled from the uniform distribution $\mathcal{U}(-1, 1)$. We drew arguments of periapse, longitudes of the ascending node, and initial mean anomalies from the uniform distribution $\mathcal{U}(0, 2\pi)$. We then calculated the sky positions ($\alpha$, $\delta$) of the primary and secondary stars. Following \citet{ElBadry2024}, we identified unresolved binaries using the criterion
\begin{equation}
    \Delta G > \frac{1}{25}\left(\frac{\rho}{\mathrm{mas}} - 200\right)
\end{equation}
where $\rho$ is the angular separation and $\Delta G$ is the magnitude difference between the secondary and primary stars ($\Delta G\,{=}\,G_2\,{-}\,G_1$). We discarded resolved binaries and unresolved binaries with a combined $G\,{>}\,19$, for which astrometric motion is unlikely to be detectable by {\it Gaia}. After all of these cuts, we were left with a sample of $43$~million unresolved binary stars within $2{,}000$~pc with $G\,{<}\,19$, in good agreement with \citet{ElBadry2024}, whose model predicted $46$~million.

\subsection{False-positive predictions}

We forecasted binary detections by fitting simulated epoch astrometry, as in Section~\ref{sec:mock_catalog}. We restricted our attention to the subset of the $43$~million binaries that were plausibly detectable: those with \snr$\,{>}\,0.5$, and $P_\mathrm{orb}$ shorter than $7$~years for DR4 and $14$~years for DR5. In this case, the \snr\ was calculated as the ratio of the amplitude of {\it photocenter} motion to the astrometric precision per data point (Equation~\ref{eqn:astrom_unc}). We calculated the amplitude of photocenter motion using
\begin{equation}
\label{eqn:photocenter_rad}
    a_0 = a \left|\frac{F_1}{F_1+F_2} - \frac{M_1}{M_1 + M_2}\right|~,
\end{equation}
where $a$ is the semi-major axis of the relative orbit, as usual, and $F_1$ and $F_2$ are the $G$-band fluxes of the primary and secondary stars. For each binary that (1) passed the \snr\ threshold, (2) satisfied the orbital period limit, and (3) had a value of $a_0$ small enough to be compatible with a ${<}\,30\,M_\mathrm{J}$ dark companion orbiting the primary star, we simulated time-series astrometry for the binary using \texttt{Gaiamock} (see Section~\ref{sec:astrom_data}). We fitted the simulated data with no-companion and one-companion models.

For a direct comparison with the planet-yield results presented in Section~\ref{sec:mock_catalog}, we applied the same detection criteria to the planet impostors: $\Delta \chi^2\,{>}\,50$, best-fit $m_p\,{<}\,13\,M_{\rm J}$, and $P_\mathrm{84th}/P_\mathrm{16th}\,{<}\,1.5$. The planet mass requirement depends on the ``apparent mass'' of the unresolved binary. That is, the mass that would be assigned to a single star with the same total $G$-band flux and $G_{BP}\,{-}\,G_{RP}$ color of the unresolved binary. To assign apparent masses to binaries, we summed the $G$, $G_{BP}$, and $G_{RP}$ fluxes of the two binary components to obtain combined apparent magnitudes, and fitted these magnitudes with MIST isochrones \citep{Choi2016}. The resulting apparent masses were generally $10$\,--\,$50$\% larger than the stars in the nearly equal-mass binaries. We used these apparent masses when inferring planet masses, and we show the apparent masses in the plots below.

We recorded a total of $1{,}200$ binary false-positive detections in DR4 and $6{,}300$ in DR5. This represents $13$\% of the total number of planet candidates (planets and unresolved binaries) in DR4 and $5$\% in DR5. Although genuine planet detections dominate overall, the fraction of false positives depends strongly on the system parameters. Figure~\ref{fig:planet_binary_hists} shows the ratio of the number of genuine planet detections to the total number of detections. Relative to the number of planet detections, impostors are more probable when the source has a low apparent mass, a faint apparent magnitude, or a close-orbiting planet. Small orbits are preferred by impostors because this results in lower-amplitude astrometric motion, allowing them to mimic planetary-scale signals without as much fine-tuning of the flux ratio (Equation~\ref{eqn:photocenter_rad}). The preference of impostors for low stellar masses has a similar explanation: at fixed orbital period, lower-mass binaries have smaller orbits. In DR5, the false positive rate for low-mass stars and faint stars will be somewhat lower than in DR4. The DR4 and DR5 false positive rates for close-orbiting planets (${\lesssim}\,1$\,AU) are expected to be similar.

\begin{figure}
\centering
\includegraphics[width=0.475\textwidth]{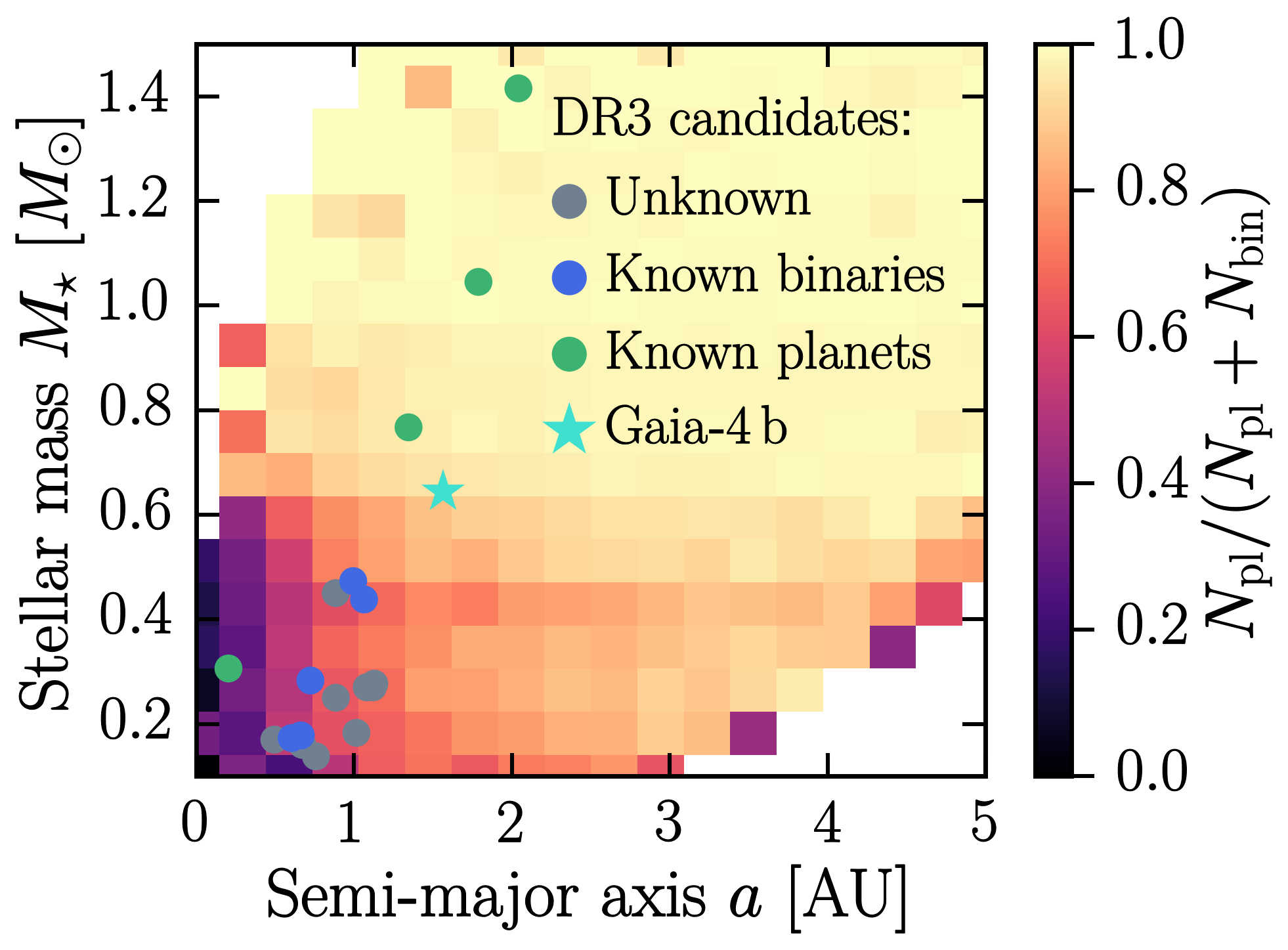}
\caption{The fraction of DR5 planet candidates that are genuine planets, instead of unresolved binaries, for different values of apparent stellar mass and semi-major axis. The points are {\it Gaia} DR3 exoplanet candidates with $m_p\,{<}\,13\,M_{\rm J}$, colored based on our current knowledge of their status. Most of the DR3 candidates are close-in giants around low-mass stars, a region of parameter space where we expect severe contamination by unresolved binaries. The confirmed planet Gaia-4\,b \citep{Stefansson2025} and most of the other previously known planets are in regions of higher purity.}
\label{fig:2D_planet_binary_hists}
\end{figure}

Most of the {\it Gaia} DR3 exoplanet candidates are close-orbiting super-Jupiters around low-mass stars, placing them in the regime where we expect contamination by binaries to be especially severe. Figure~\ref{fig:2D_planet_binary_hists} shows the fraction of DR5 detections that are genuine planets for different combinations of apparent stellar mass and semi-major axis. The {\it Gaia} DR3 exoplanet candidates with $m_p\,{<}\,13\,M_{\rm J}$ from the ``orbital'' catalog and previously known planets are also plotted \citep{GaiaCollaboration2023}. Colors indicate the current status of the candidates (see the follow-up efforts of \citealt{Marcussen2023, Stefansson2025}). Most of the DR3 planet candidates, including all of the known impostors, have $M_\star\,{\lesssim}\,0.6\,M_\odot$ and $a\,{\lesssim}\,1$\,AU, where we expect the false positive rate to be about $50$\% in both DR4 and DR5. The shorter baseline of DR3 and the complex selection process used to create the catalog might have further increased the false positive fraction. The previously known planets (green) lie in regions with lower expected false positive rates, except for GJ\,876\,b \citep{Delfosse1998, Marcy1998}, which features a short orbital period ($61$~days) and a low-mass host star that is unusually nearby ($4.7$~pc). The genuine planet Gaia-4\,b has a wider orbit and a more massive host star than the candidates that turned out to be binaries, as one might have expected based on our simulations.

\subsection{Identifying planet-impostor binaries}

In some cases, planet-impostor binaries might be identifiable from {\it Gaia} data alone, without the need for ground-based spectroscopic follow-up. Spectra provided by the {\it Gaia} RVS can identify binaries by observing two distinct sets of absorption lines \citep{GaiaCollaboration2023} or anomalously broad lines due to the components of the binary \citep[e.g.,][]{Hadad2025}. However, the resolution of RVS spectra is modest ($R\,{\approx}\,11{,}500$) and its precision deteriorates for faint stars \citep{Katz2023}. In DR3, the line broadening parameter (\texttt{vbroad}) was only provided for stars brighter than $G\,{=}\,12$, and its accuracy was found to degrade sharply for $G\,{>}\,10$ \citep{Fremat2023}. We expect the contamination due to planet impostors to be minor for sources with $G\,{<}\,12$ (${\lesssim}\,10$\%; see Figure~\ref{fig:planet_binary_hists}). Thus, it seems likely that the {\it Gaia} RVS spectra will be of limited utility for identifying planet impostors, unless the spectra for faint stars improve significantly in the coming data releases.

{\it Gaia} colors provide another possible means for identifying planet-impostor binaries. Unresolved twin binaries possess a color that is similar to the companion stars, but a luminosity that is twice as large, which can make them outliers on a color-magnitude diagram. Because stars with detectable planets and planet-impostor binaries are a special subset of single and binary systems, it is not obvious how many binaries can be discarded based on their colors and luminosities. This question can be tackled using our mock planet catalog (Section~\ref{sec:mock_catalog}) and planet impostor catalog (Section~\ref{sec:binary_stars}). For a meaningful comparison with simulated binaries, {\it Gaia} magnitudes must be corrected for extinction. We did so using the $G$, $G_{BP}$, and $G_{RP}$ extinctions reported in the DR3 source catalog, discarding sources which lack {\it Gaia}-reported extinctions. To remove possible binaries in our DR5 mock catalog, we also discarded sources that display excess astrometric noise, as quantified with the \texttt{RUWE} parameter (see, e.g., \citealt{Belokurov2020}). Specifically, we discarded sources from the catalog for which \texttt{RUWE}\,${<}\,1.2$. We note that planets can also cause singles to have inflated \texttt{RUWE} values, but inadvertently discarding some singles would not meaningfully affect this comparison. These cuts left us with $75{,}000$ planet-hosting (probable) singles with extinction-corrected {\it Gaia} colors and $6{,}300$ unresolved binaries with simulated colors.

\begin{figure}
\centering
\includegraphics[width=0.475\textwidth]{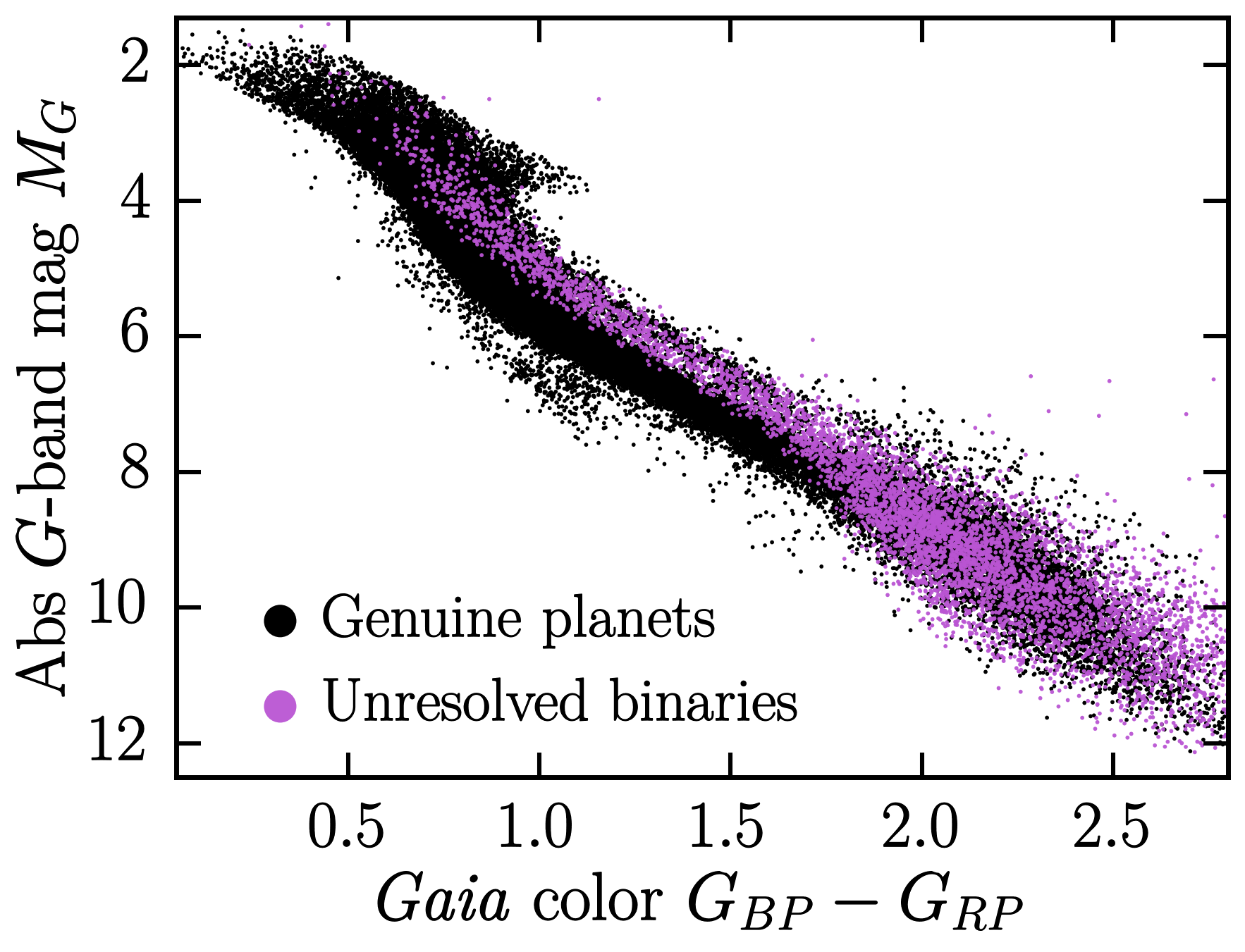}
\caption{Color-magnitude diagram for {\it Gaia} stars in our mock DR5 exoplanet catalog (black) and simulated unresolved binaries with planet-like astrometric motion (purple). {\it Gaia} magnitudes were corrected using the extinctions reported in the DR3 source catalog. In an effort to remove binaries, we discarded stars with \texttt{RUWE}\,${>}\,1.2$. Most unresolved binaries cannot be excluded based only on color and luminosity. At the cost of discarding $5$\% of genuine planets, about $20$\% of unresolved binaries can be removed, mostly those in the range $1\,{\lesssim}\,G_{BP}\,{-}\,G_{RP}\,{\lesssim}\,2$.}
\label{fig:CMD_binaries}
\end{figure}

Figure~\ref{fig:CMD_binaries} compares the colors and magnitudes of the stars hosting detected planets and the planet-impostor binaries. For sources with $1\,{\lesssim}\,G_{BP}\,{-}\,G_{RP}\,{\lesssim}\,2$, unresolved binaries are somewhat brighter than stars with planets, although the difference is modest. Discarding $20$\% of the planet impostors using a color-dependent luminosity threshold also removes about $5$\% of singles with planets. As such, we expect that a fairly small fraction of unresolved binaries will be readily identifiable based on their {\it Gaia} colors and luminosities alone.

The colors and luminosities of our simulated stars could differ systematically from those of real {\it Gaia} stars. To check whether it is safe to compare {\it Gaia} stars with our simulated binaries, we repeated this analysis with simulated singles. Specifically, we took the $680$~million single stars generated using \texttt{COSMIC}, as described in Section~\ref{sec:sim_binaries}. Then, we assigned $G$, $G_{BP}$, and $G_{RP}$ magnitudes using the MIST isochrones, restricted our attention to observationally relevant stars (as in Section~\ref{sec:obs_rel_stars}), assigned planets according to our planet occurrence model (Section~\ref{sec:planet_occ}), and tallied planet detections using our simple DR5 detection criteria (\snr\,${>}\,1.0$ and $P_\mathrm{orb}\,{<}\,9.5$\,years). We found the colors and absolute magnitudes of these simulated stars to resemble those of real stars in our mock catalogs, reinforcing the conclusions above.

\section{Discussion}
\label{sec:discussion}

Below, we compare our results with earlier works, discuss potential caveats, and summarize our findings.

\subsection{Comparison with earlier projections}
\label{sec:comparisons}

As mentioned in Section~\ref{sec:intro}, the most comprehensive prior estimates for {\it Gaia}'s planet yield were performed by \citet{Perryman2014}. Whereas we predict that {\it Gaia} will discover $7{,}500\,{\pm}\,2{,}100$ planets in DR4 and $120{,}000\,{\pm}\,22{,}000$ in DR5, \citet{Perryman2014} predicted there would be $21{,}000\,{\pm}\,6{,}000$ in DR4 and $70{,}000\,{\pm}\,20{,}000$ in DR5, respectively. There are many possible reasons for the discrepancies between the predictions, but an especially notable difference is related to the detection criteria. We recorded detections whenever $\Delta \chi^2\,{>}\,50$, the orbital period was at least modestly constrained ($P_\mathrm{84th}/P_\mathrm{16th}\,{<}\,1.5$), and the planet's best-fit mass was below $13~M_\mathrm{J}$. The uncertainty in our projection reflects the uncertainty in the planet occurrence function. In contrast, \citet{Perryman2014} required detections to satisfy $P_\mathrm{orb}\,{<}\,6$\,years and $m_p\,{<}\,15\,M_\mathrm{J}$. The quoted result of $21{,}000\,{\pm}\,6{,}000$ was based on the spread in the number of detections they recorded when considering $\Delta \chi^2$ thresholds between $30$ and $50$, which they labeled as ``marginal'' and ``reliable'' thresholds, respectively. Because the amplitude of the astrometric signal grows sharply with $P_\mathrm{orb}$, their sample is dominated by planets with $P_\mathrm{orb}\,{\approx}\,6$\,years, which typically have poorly constrained orbits. This also explains why planets in their DR4 simulation extend out to larger orbital separations than planets in our DR4 catalog ($5$~AU versus $3$~AU; see their Figure~2 and our Figure~\ref{fig:mock_catalog}). For a more direct comparison, we tried dropping the $P_\mathrm{84th}/P_\mathrm{16th}\,{<}\,1.5$ condition and instead using a period threshold of $P_\mathrm{orb}\,{<}\,6$\,years. This increased our predicted number of detections to $16{,}000$. Extrapolating our results to allow for planet masses as large as $15\,M_\mathrm{J}$ and $\Delta \chi^2$ values as small as $40$ (the midpoint of the range in \citealt{Perryman2014}) resulted in about $26{,}000$ detections. The agreement with \citet{Perryman2014} is good but probably somewhat coincidental, because they adopted astrometric uncertainties that were about $50$\% lower for bright stars. The differing stellar populations and planet occurrence models may have compensated for this difference.

Our simulations led us to regard $\Delta \chi^2\,{>}\,30$, the ``marginal'' detection threshold from \citet{Perryman2014} and \citet{Holl2022}, as too weak to be a useful detection threshold for most purposes. Invoking a model with seven additional free parameters inevitably results in a better fit to the astrometric data, with a typical reduction in $\chi^2$ of $20$ for simulated DR4 data even when there was no companion (see Section~\ref{sec:SNR_chi2_delta}). Adopting a threshold of $\Delta \chi^2\,{>}\,30$ would probably lead to an intolerably high level of statistical false positives (``flukes''). In the experiments described in Section~\ref{sec:SNR_chi2_delta}, $5$\% of the sources {\it without} companions had $\Delta \chi^2\,{>}\,30$ purely by chance. Spurious detections around $5$\% of stars would overwhelm the number of genuine detections. Furthermore, for our simulated DR4 data, $\Delta \chi^2\,{=}\,30$ corresponds to $\Delta \mathrm{BIC}\,{\approx}\,-1$, indicating insufficient statistical evidence for a companion. A more reasonable minimum detection threshold is $\Delta \chi^2\,{>}\,50$. This corresponds to $\Delta \mathrm{BIC}\,{\gtrsim}\,19$, indicating a meaningful (but still modest) preference for the one-companion model.

Based on orbit-fitting experiments, \citet{Casertano2008} reported that systems with \snr\,${=}\,6$ typically had orbital parameters and planet masses measured to within $15$\,--$20$\% with {\it Gaia} DR4 data. Our results agree for all quantities except the eccentricity. Based on Equations~\ref{eqn:DR4_Porb_SNR}\,--\,\ref{eqn:DR4_ecc_SNR} when \snr\,${=}\,6$, we expect the median relative error to be about $0.8$\% on $P_\mathrm{orb}$, $3$\% on $m_p$, $3$\% on $i$, and $40$\% on $e$, with a large spread. Indeed, $84$\% of systems in our DR4 orbit fitting experiments with \snr\,${\approx}\,6$ had relative errors of less than $15$\% on $P_\mathrm{orb}$, $m_p$, and $i$. \citet{Perryman2014}'s orbit constraint predictions were more optimistic. They reported that cases with $\Delta \chi^2\,{>}\,100$ in DR4 typically provided orbital parameters ``determined to $10$\% or better.'' We found this to be true for the orbital period, but not for other parameters of interest. According to Equation~\ref{eqn:DR4_chi2_SNR}, $\Delta \chi^2\,{=}\,100$ corresponds to \snr\,${\approx}\,2$. Equations~\ref{eqn:DR4_Porb_SNR}\,--\,\ref{eqn:DR4_ecc_SNR} were based on a fit to systems with \snr\,${>}\,3$, and are unreliable for lower values of \snr. $84$\% of systems in our DR4 orbit fitting experiments with \snr\,${\approx}\,2$ had $\delta P_\mathrm{orb}\,{<}\,0.1$, but only $28$\% and $46$\% had $\delta m_p\,{<}\,0.1$ and $\delta i\,{<}\,0.1$, respectively. For the errors in $m_p$ and $i$ to be below $10$\% for more than $84$\% of systems, \snr\ must be at least $6$ ($\Delta \chi^2\,{\approx}\,800$).

Giant planets orbiting M-dwarf stars are a population of particular interest because of their potential to test the core-accretion theory \citep[e.g.,][]{Ida2005, Miguel2020, Burn2021}. \citet{Sozzetti2014} and \citet{Perryman2014} predicted that there would be ${\sim}\,2{,}600$ and ${\sim}\,1{,}000$ DR4 detections around M-dwarfs, respectively, all of which are around nearby stars ($r\,{<}\,100$\,pc). Our forecast for DR4 is comparable. Our mock exoplanet catalogs include $2{,}200\,{\pm}\,630$ detections around M-dwarfs in DR4 and $20{,}000\,{\pm}\,3{,}800$ in DR5. Here, we have adopted the same relative uncertainties as on the total detection numbers, but in reality, the planet occurrence function is somewhat more uncertain for low-mass stars (see Section~\ref{sec:unc_planet_occ}).

\subsection{Potential Complications}
\label{sec:complications}

In the preceding analyses, we neglected several complications that will arise in the real astrometric planet detection process. Here, we discuss the potential impact of multiple-planet systems, planets that exist in stellar binaries, and misclassified brown dwarfs.

Many stars observed by {\it Gaia} -- perhaps the majority of stars -- will host multiple planets. We have not investigated how many stars will host more than one {\it Gaia}-detectable planet (i.e., multiple super-Jupiters on several-AU orbits), how challenging it will be to distinguish these systems from single-planet systems, and how accurately the planet parameters can be recovered. A few multi-super-Jupiter systems have been discovered using the Doppler method (e.g., HD\,125612, HD\,183263, and HD\,203387; \citealt{Wright2009, LoCurto2010, Feng2022}). In each case, the super-Jupiters' orbits are widely spaced ($\Delta a\,{\gtrsim}\,2.5$\,AU), resulting in fairly distinct orbital periods and amplitudes of their astrometric signals that facilitate clean retrievals. In fact, the requirement for long-term dynamical stability leads to a firm expectation that super-Jupiters will be relatively widely spaced. For a $5$~$M_{\rm J}$ planet on a $3$~AU orbit around a $0.8$~$M_\odot$ star to be accompanied by a $5$~$M_{\rm J}$ outer companion on a Hill-stable orbit (i.e., $\Delta a\,{>}\,2\sqrt{3}R_H$; \citealt{Gladman1993}), it must have a semi-major axis of at least $5.3$~AU. The spacings must be even larger if the planets have non-zero orbital eccentricities. Because {\it Gaia} is sensitive to planets within a relatively narrow range of orbital separations (see Figure~\ref{fig:mock_catalog}), it may prove unlikely for stars to host more than one {\it Gaia}-detectable planet.

In addition to mimicking planet-like astrometric motion, stellar binaries can produce other complications. Many planet-hosting stars will be members of binaries. In some cases, the motion induced by the planet and the binary orbit will both be detectable, leading to more complex astrometric signals. Fortunately, the members of the binary cannot be too close without making the system dynamically unstable. According to the empirical stability threshold of \citet{Holman&Wiegert1999}, for an $0.8$~$M_\odot$ star to host a (massless) planet on a $3$~AU orbit and a $0.1$~$M_\odot$ binary companion on a circular orbit, the binary separation must be ${\gtrsim}\,7$\,AU. This corresponds to an orbital period of ${\gtrsim}\,20$\,years, which is longer than the $10$-year duration of the {\it Gaia} mission and several times longer than the $5.8$~year period of the planet. More massive or eccentric binary companions would require longer orbital periods to permit dynamical stability. Thus, the main effect of binary companions will be to induce long-term accelerations that could complicate orbit fitting. Cross-matching with {\it Hipparcos} data may provide additional information about the orbits of some long-period binaries (see \citealt{Brandt2018, Brandt2021}). A separate concern is the influence of binaries on the occurrence rate of super-Jupiters, which is currently poorly constrained. Super-Jupiters could be more likely, or less likely, to form in binary systems than around single stars. Although this makes predictions more difficult, it also means that {\it Gaia} might teach us about how planet occurrence depends on binarity.

Brown dwarfs provide another source of false positives, or at least definitional ambiguity, depending on one's perspective. The top-heavy mass distribution of detected planets (Figure~\ref{fig:mock_catalog}) leads one to expect that there will be more brown dwarf detections than planet detections. Some brown dwarfs will be misclassified as giant planet candidates, especially in the low-SNR regime (\snr\,${\lesssim}\,4$ in DR4 and \snr\,${\lesssim}\,2.5$ in DR5), where companion masses will be poorly constrained. Using the {\it Gaia} DR2 catalog and the Besançon stellar population model \citep{Robin2003}, \citet{Holl2022} predicted that {\it Gaia} astrometry would detect ${\sim}\,30{,}000$ brown dwarfs of mass $10$\,--\,$80\,M_\mathrm{J}$ in DR4 and ${\sim}\,50{,}000$ in DR5. These projections are relatively uncertain and cannot be compared directly with ours, but solidify the expectation that {\it Gaia} will detect tens of thousands of brown dwarfs. RV follow-up will be important to establish the planetary nature of the {\it Gaia} detections.

\subsection{Summary}
\label{sec:summar}

The first release of {\it Gaia}'s time-series astrometry promises to open the floodgates for astrometric planet detection. With an updated understanding of giant-planet occurrence statistics, the characteristics of nearby stars, and {\it Gaia}'s astrometric precision, we have forecasted {\it Gaia}'s exoplanet detections for Data Release 4 (scheduled for December 2026) and the eventual end-of-mission Data Release 5 (not before 2030). Our conclusions are summarized below.
\begin{itemize}

    \item {\it Gaia} can detect super-Jupiters ($m_p\,{\gtrsim}\,3M_\mathrm{J}$) in several-AU orbits around stars within several hundred parsecs. The search volume is much smaller for Jupiter-mass planets and planets within an AU. Planets around massive stars ($M_\star\,{\gtrsim}\,1.5M_\odot$) are only detectable for bright hosts ($G\,{\lesssim}\,11$), making such detections rare. We provide predictions for {\it Gaia}'s search radius and limiting magnitude (Figure~\ref{fig:rmax_Gmax}) and the number of searchable stars (Figure~\ref{fig:num_searchable}) to help understand {\it Gaia}'s sensitivity.

    \item We predict that {\it Gaia} will discover $7{,}500\,{\pm}\,2{,}100$ planets in DR4 and $120{,}000\,{\pm}\,22{,}000$ planets in DR5. The dominant known source of uncertainty for this prediction comes from the uncertainty in the planet occurrence function. The number of candidates that are ultimately identified will depend sensitively on the strictness of the detection criteria. We adopted the following detection criteria: the one-companion model is favored over the no-companion model by $\Delta \chi^2\,{>}\,50$, the best-fit planet mass is below $13\,M_\mathrm{J}$, and the orbital period is moderately constrained ($P_\mathrm{84th}/P_\mathrm{16th}\,{<}\,1.5$). Figure~\ref{fig:mock_catalog} shows how the detection numbers depend on these choices.

    \item Most of the detected planets will be super-Jupiters ($3$\,--\,$13\,M_\mathrm{J}$) in several-AU orbits ($1$\,--$3$\,AU for DR4; $2$\,--$5$\,AU for DR5) around GKM-type stars ($0.4$\,--\,$1.2\,M_\odot$). Host stars will be located within ${\sim}\,250$\,pc for DR4 and ${\sim}\,500$\,pc for DR5, with $G$-band magnitudes between $9$ and $16$ (see Figure~\ref{fig:mock_catalog}). Most detections will be made with modest statistical significance and will not provide accurate measurements of planet mass and inclination. The expected number of planets for which masses and orbital periods will be measured to within $20$\% is $1{,}900\,{\pm}\,540$ for DR4 and $38{,}000\,{\pm}\,7{,}300$ for DR5.

    \item {\it Gaia}'s ability to constrain orbital elements and planet masses can be approximated as a simple function of \snr, the per-point astrometric SNR (Equation~\ref{eqn:SNR_criterion}). Based on simulated {\it Gaia} astrometry, we showed that the relative error in the orbital period, planet mass, and inclination scale inversely with \snr. We reported fits to the data for DR4 (Equations~\ref{eqn:DR4_chi2_SNR}\,--\,\ref{eqn:DR4_ecc_SNR}) and DR5 (Equations~\ref{eqn:DR5_chi2_SNR}\,--\,\ref{eqn:DR5_ecc_SNR}) to help forecast {\it Gaia}'s measurement precision.

     \item Nearly equal-mass unresolved binaries with low-amplitude photocentric motion will provide a source of false positives. However, we do not expect them to overwhelm genuine planet detections. Based on the demographics of main-sequence binaries, we predict that planet detections will outnumber detections of planet-impactor binaries by a factor of ${\sim}\,6$ in DR4 and ${\sim}\,19$ in DR5. The contamination due to binaries is much worse (about $50$\%) for close-in planets ($a\,{<}\,1$\,AU) around low-mass stars ($M_\star\,{<}\,0.6\,M_\odot$), possibly explaining why many exoplanet candidates from DR3 have turned out to be binaries. {\it Gaia} colors and RVs will probably be of limited utility for identifying planet impostors.

    \item To facilitate community preparation, we have released our mock {\it Gaia} DR4 and DR5 exoplanet catalogs, planet-impostor binary catalogs, and code to reproduce our results.\footnote{\url{https://github.com/CalebLammers/GaiaForecasts}. A frozen version is available on Zenodo: \dataset[doi: 10.17909/T9XG63]{\doi{10.5281/zenodo.17568375}}.}
    
\end{itemize}

\leavevmode

\section{Acknowledgments}
\label{sec:acknowledgments}

We thank Simon Albrecht, Timothy Brandt, Kareem El-Badry, and Benjamin Fulton for helpful discussions. C.L.\ acknowledges support from a Natural Sciences and Engineering Research Council of Canada (NSERC) Postgraduate Scholarship.

We are pleased to acknowledge that the work reported in this paper was substantially performed using the Princeton Research Computing resources at Princeton University, which is a consortium of groups led by the Princeton Institute for Computational Science and Engineering (PICSciE) and the Office of Information Technology’s Research Computing.

\appendix

\section{Astrometric planet masses}
\label{sec:cubic_mass_eqn}
A dark stellar companion of mass $m_p$ and orbital radius $a$ will induce reflex motion about the system's barycenter with radius
\begin{equation}
    a_\star = a\left(\frac{m_p}{M_\star + m_p}\right)~.
\end{equation}
In terms of the orbital period, $P_\mathrm{orb}$, this equation becomes
\begin{equation}
    m_p = a_\star \left(\frac{4 \pi^2}{G P_\mathrm{orb}^2}\right)^{\!\!1/3} \,(M_\star + m_p)^{2/3}\,~,
\end{equation}
which can be rewritten as
\begin{equation}
\label{eqn:cubic_q}
\left( \frac{a_M}{a_\star} \right)^{\!\!3}\,q^3 - q^2 - 2q - 1 = 0~,
\end{equation}
where $q\equiv m_p/M_\star$ and
\begin{equation}
a_M \equiv 
\left(\frac{GM_\star P_\mathrm{orb}^2}{4\pi^2} \right)^{\!\!1/3}.
\end{equation}
The left side of Equation~\ref{eqn:cubic_q} is negative when $q\,{=}\,0$ and increases indefinitely as $q\,{\rightarrow}\,\infty$, guaranteeing the existence of a positive $q$ root. Additionally, since there is only one sign change between coefficients, it follows from Descartes' rule of signs that the positive $q$ root is unique.

Under the assumption that $q\,{\ll}\,1$, the cubic relation in Equation~\ref{eqn:cubic_q} simplifies to $q\,{\approx}\,a_\star/a_M$. This provides a suitable approximation for low-\snr\ systems, but for systems with \snr\,${\gtrsim}\,10$, we found this approximation to limit the accuracy with which planet masses can be measured (see Section~\ref{sec:SNR_chi2_delta}). As such, we determined planet masses in our orbit fitting experiments by measuring $a_\star$, then finding the positive $q$ root of Equation~\ref{eqn:cubic_q}. For a related formulation of the astrometric companion problem, with an emphasis on dormant black holes, see \citet{Shahaf2019}.

\bibliography{refs}{}
\bibliographystyle{aasjournal}

\end{document}